%% file: paper.tex
\documentclass{sig-alternate}
\pdfoutput=1
\usepackage[utf8]{inputenc}
\usepackage{url}
\usepackage[scaled]{helvet}
\usepackage[T1]{fontenc}
\usepackage{graphicx}
\usepackage[american]{babel}
\usepackage{xspace}
\usepackage{fancyvrb}
\usepackage{amsmath}
\usepackage{amsfonts}
\usepackage{tikz}
\usepackage{rotating}
\usepackage{booktabs}
\usepackage{tabularx}
\usepackage{color,pdfcolmk}
\usepackage[normalem]{ulem}
\usepackage{array}
\usepackage{subfig}
\usepackage[pdftex]{hyperref}

\usepackage[pass]{geometry}

\usepackage{microtype}
\usepackage[font=bf,belowskip=-4pt,aboveskip=2pt]{caption}
\DeclareCaptionType{copyrightbox}

\newcommand{\mypar}[1]{\smallskip\textbf{#1}.}

\usepackage{pgfplots}
\usepackage{pgfplotstable}
\pgfplotsset{compat=newest}
\usetikzlibrary{decorations.pathmorphing}
\pgfkeys{/pgfplots/ComparisonPlot/.style={height=7cm, width=0.50\paperwidth,scale only axis}}
\pgfkeys{/pgfplots/NumberPlotLarge/.style={width=0.40\paperwidth,scale only axis}}
\pgfkeys{/pgfplots/NumberPlotSmall/.style={height=5cm, width=0.30\paperwidth,scale only axis}}

\makeatletter
\tikzset{nomorepostaction/.code=\let\tikz@postactions\pgfutil@empty}
\tikzstyle{every node}=[font=\scriptsize]
\makeatother

\makeatletter
\newcommand{\resetstackedplots}{
  \makeatletter
  \pgfplots@stacked@isfirstplottrue
  \makeatother
  \addplot [forget plot,draw=none] coordinates{(NoStim,0) (Monkey,0) (Copper,0) (Our,0)};
}
\makeatother

\pgfdeclarelayer{background}
\pgfdeclarelayer{foreground}
\pgfsetlayers{background,main,foreground}
\usetikzlibrary{shadows}
\usetikzlibrary{fit}
\usetikzlibrary{patterns}
\usetikzlibrary{calc}
\usetikzlibrary{matrix}
\usetikzlibrary{arrows}
\usetikzlibrary{calc}
\usetikzlibrary{fit}
\usetikzlibrary{patterns}
\usetikzlibrary{shadows}
\usetikzlibrary{shapes.symbols}
\usetikzlibrary{shapes.geometric}
\usetikzlibrary{shapes.multipart}
\usetikzlibrary{shapes.misc}
\usetikzlibrary{shapes.arrows}
\usetikzlibrary{decorations.pathmorphing}
\usetikzlibrary{shapes,decorations,shadows}
\usetikzlibrary{decorations.pathmorphing}
\usetikzlibrary{decorations.shapes}
\usetikzlibrary{decorations.text}
\usetikzlibrary{decorations.pathreplacing}
\usetikzlibrary{fit,arrows,calc,positioning}
\pgfdeclarelayer{foreground}
\pgfdeclarelayer{background}
\pgfdeclarelayer{middle}

\newcommand{\therealname}{PuppetDroid}

\newcommand{\copperdroid}{CopperDroid\xspace}
\newcommand{\strace}{stimulation trace\xspace}
\newcommand{\straces}{stimulation traces\xspace}

\newcommand{\thesystemtitle}{\therealname\xspace}
\newcommand{\thesystem}{\textsf{\thesystemtitle}\xspace}
\newcommand{\thetitle}{A User-Centric UI Exerciser for Automatic
  Dynamic Analysis of Similar Android Applications}
\newcommand{\mail}[1]{\href{mailto:#1}{#1}}

\hypersetup{
    bookmarks=true,
    pdffitwindow=false,
    pdftitle={\thesystemtitle: \thetitle},
    pdfauthor={Andrea Gianazza, Federico Maggi, Aristide Fattori, Lorenzo Cavallaro, Stefano Zanero},
    pdfsubject={Subject},
    pdfcreator={Federico MaggiAndrea Gianazza, Federico Maggi, Aristide Fattori, Lorenzo Cavallaro, Stefano Zanero},
    pdfkeywords={Android} {Testing} {Dynamic Analysis},
    colorlinks=false,
}

\begin{document}

\title{\thesystemtitle: \thetitle}

\author{%
  \begin{tabular}{ccc}
    \begin{minipage}{0.3\linewidth}
      \centering\large\sffamily
      Andrea Gianazza\\
      Politecnico di Milano\\
      \mail{andrea.gianazza@mail.polimi.it}
    \end{minipage}
    &
    \begin{minipage}{0.3\linewidth}
      \centering\large\sffamily
      Federico Maggi\\
      Politecnico di Milano\\
      \mail{federico.maggi@polimi.it}
    \end{minipage}
  \end{tabular}\\[1cm]
  \begin{tabular}{ccc}
    \begin{minipage}{0.33\linewidth}
      \centering\large\sffamily
      Aristide Fattori\\
      Univ. degli Studi di Milano\\
      \mail{joystick@security.di.unimi.it}
    \end{minipage}
    &
    \begin{minipage}{0.33\linewidth}
      \centering\large\sffamily
      Lorenzo Cavallaro\\
      Royal Holloway Univ. of London\\
      \mail{lorenzo.cavallaro@rhul.ac.uk}
    \end{minipage}
    &
    \begin{minipage}{0.33\linewidth}
      \centering\large\sffamily
      Stefano Zanero\\
      Politecnico di Milano\\
      \mail{stefano.zanero@polimi.it}
    \end{minipage}
  \end{tabular}}

\maketitle

\begin{abstract}
\input{abstract}
\end{abstract}

\input{introduction}
\input{background}
\input{core}
\input{evaluation}
\input{future}
\input{related}
\input{conclusion}

\subsection*{Acknowledgements}
This research has been partially funded under the EPSRC Grant
Agreement EP/L022710/1 and by the FP7 project SysSec funded by the EU
Commission under grant agreement no. 257007.

\bibliographystyle{abbrv}
\bibliography{biblio}

\input{appendix}

\end{document}

%% file: abstract.tex
Popularity and complexity of malicious mobile applications are rising,
making their analysis difficult and labor intensive. Mobile application analysis
is indeed inherently different from desktop application analysis: In the latter,
the interaction of the user (i.e., victim) is crucial for the malware to
correctly expose all its malicious behaviors.

We propose a novel approach to analyze (malicious) mobile
applications. The goal is to exercise the user interface (UI) of an Android
application  to effectively trigger malicious behaviors,
automatically. Our key intuition is to record and reproduce the
UI interactions of a potential victim of the malware, so as to stimulate the
relevant behaviors during dynamic analysis. To make our approach
scale, we automatically re-execute the recorded UI interactions on apps that are
similar to the original ones. These characteristics make our system
orthogonal and complementary to current dynamic analysis and UI-exercising
approaches.

We developed our approach and experimentally shown that our stimulation
 allows to reach a higher code coverage than automatic UI exercisers, so
to unveil interesting malicious behaviors that are not exposed
when using other approaches.

Our approach is also suitable for crowdsourcing scenarios, which would push
further the collection of new stimulation traces. This can potentially change
the way we conduct dynamic analysis of (mobile) applications, from fully
automatic only, to user-centric and collaborative too.


%% file: introduction.tex
\section{Introduction}
The popularity of Android devices~\cite{idc_android_market_share} make them a
very attractive target~\cite{trendmicro-q3-2012,eset2012}. Although a wide range
of threats have been spotted in the wild, the typical vector is a malicious app
distributed through official or unofficial markets. Once installed, these apps
perform actions without the victim's consent, ending up in financial loss or
information stealing. Security companies,
researchers~\cite{Zhou_dissecting_android_malware}, and practitioners agree that
there is an increasing trend, which indicates that cyber criminals consider
malicious apps a viable
business\footnote{\url{http://cwonline.computerworld.com/t/8652955/807570490/619941/0/}}.

To develop detection and defensive countermeasures, understanding how Android
malware works is essential. After having ported the main program-analysis
techniques (e.g., dynamic analysis, static analysis, symbolic execution) to the
Android platform, we are now hitting their well-known limitations: Dynamic
analysis has scarce code coverage, static analysis is hindered by obfuscation,
and symbolic execution is resource intensive. Dynamic analysis of mobile
applications is particularly difficult because of the highly interactive nature
of mobile user interfaces (UI). A malicious Android application that is not
properly exercised may not expose its malicious behavior at all.

As overviewed in Section~\ref{sec:background} and \ref{sec:related-work},
state-of-the-art dynamic analysis approaches rely either on program analysis
(e.g., backward slicing) or stress-test tools (e.g., Monkey) to increase code
coverage. However, none of the current approaches take into account how the UI
is exercised by end users. To take dynamic analysis one step further, we believe
that a radically different approach should be pursued. In this paper we propose
a new approach to exercise the UI of an Android application that changes the way
malware analysis experiments are currently conducted, and effectively stimulate
potentially malicious behaviors.

The key intuition is to leverage human-driven UI exercising. We first show that
human users are more effective at exercising an application than automatic tools
because they understand the semantic of the UI elements, and can exercise the
application accordingly. However, the number of human resources required would
be unreasonably large. Therefore, to downsize this blocking requirement, we
leverage a second key observation: many malware samples are similar to each
other, for two reasons. First, both AV vendors~\cite{eset2012,trendmicro2012}
and research work showed that malware authors repackage existing malware samples
rather than creating brand-new
families~\cite{2013_maggi_valdi_zanero_andrototal,droidchameleon,Fedler:2013wl}. As
a result, there exist many instances of the same variant. Secondly, miscreants
hide malware inside well-known, paid apps and distribute them free of charge to
fool unaware users~\cite{VC:CODASPY13,droidmoss,hanna12:_juxtap}. Therefore, it
is common to find the same benign application used to hide many malicious
payloads. We leverage these two sources of similarity to make our approach
scale: Our idea is to record a trace of the human-driven UI stimulation, which
we name \emph{\strace}, performed during a test, and then leverage the
UI similarity to automatically re-execute this trace on applications similar to
one originally tested by the user. In this way, if at least one user in our
system succeeds in manually stimulating a malicious behavior in a malware (or in
a benign application that hides malware), it is quite likely that, by re-using
the same \strace on similar applications, we can exercise similar (malicious)
behaviors.

We design and implement \thesystem, an Android environment that supports both
\textit{manual application testing} (through a physical device or an emulator),
to collect new \straces, and \textit{automatic application exercising}, which
cleverly leverages previously recorded UI \straces. \thesystem relies on the
screenshots of an app to find similar apps, indexed using a fast metric known as
perceptual hashing.

To evaluate \thesystem, first, we experimentally verified that manual exercising
allows to stimulate malicious behaviors better than automatic
techniques. Second, we validated our approach on 7,000 applications and found
out that it can stimulate 12--24\% more behaviors than state-of-the-art
techniques. Interestingly, our system is able to unveil those corner behaviors
that are difficult to exercise with a fully automatic approach (i.e., download
an APK and execute it). Then, we show that our UI similarity technique is
precise, scales, and has modest resource requirements. In summary:
\begin{itemize}\itemsep0em
\item we propose a novel and orthogonal approach to exercise more behaviors
  during dynamic analysis of (malicious) mobile applications. Our approach is
  the first that takes the end users into play.
\item We propose an original method to automatically exercise the UI of an
  unknown application re-using UI stimulation traces obtained from previously
  analyzed applications that present a similar layout.
\item We implemented and evaluated our approach to demonstrate its feasibility
  and, more importantly, its effectiveness. Remarkably, we manually verified the
  outcome of each experiment.
\end{itemize}


%% file: background.tex
\section{Background and Motivation}
\label{sec:background}
Many approaches have been proposed to analyze applications with the final goal
of designing effective detection
criteria~\cite{droidscope,andrubis_blog,copperdroid,enck2010taintdroid,droidbox,droidranger,droidmoss,riskranker,oberheide:bouncer}. To
this end, program-analysis techniques used for traditional malware have been
ported to Android (e.g., dynamic analysis, static analysis, taint tracking,
symbolic execution), with their well-known, symmetric pros and cons. Static
approaches can be hindered by obfuscated code, repackaging or dynamic payloads,
two techniques widely used by modern malware. Symbolic execution
(e.g.,~\cite{Yang:2013:AAS:2541806.2516676}) is promising yet resource
intensive.

In spite of its efficiency and semantic richness, the main inherent limitation
of dynamic analysis is its inability to obtain satisfactory code coverage:
Dynamic analysis can examine the actions performed in an execution path only if
that path is actually explored. This problem is particularly concerning because
if a malware sample is not properly exercised, it may not expose its malicious
behavior at all. Exercising mobile applications in a proper way, however, is not
trivial, because of the highly interactive UI, which makes automatic exercising
even harder than in conventional desktop scenarios.

State-of-the-art dynamic analysis approaches (e.g.,~\cite{copperdroid})
incorporate automatic code-exercising and stimulation techniques. Other
approaches leverage stress-test tools (e.g., Monkey~\cite{androidDoc_monkey}) or
program analysis (e.g., SmartDroid~\cite{smartdroid},
ActEVE~\cite{anand2012automated}). Stress-test tools rely on pseudo-random
generation of UI input events, which is simple to implement, but rather
ineffective, since randomly stimulating UI elements displayed on the screen can
hardly reproduce the typical usage of users. Approaches such as SmartDroid
leverage static analysis to reconstruct the semantic of UI elements on the
screen, and to find execution paths that expose malicious
behaviors. Unfortunately, static analysis is ineffective against obfuscated
samples, and the research tools will need a major upgrade with the
introduction of new Android 4.4 runtime.

From our overview of the state of the art and related work in
Section~\ref{sec:related-work}, we notice that previous work does not consider
how the UI is exercised by a user. Interestingly, our experiments in
Section~\ref{sec:human} confirm our intuition that a human user is able to
exercise certain behaviors that the state of the art code stimulation
approach~\cite{copperdroid} fail to unveil.

Given the above motivations, we conclude that to take dynamic analysis of
Android applications a step further, we need an orthogonal approach to stimulate
the UI.


%% file: core.tex
\section{Goals and Approach Overview}
\label{sec:puppetdroid}
Our first goal is to provide a sandboxed environment to safely perform manual
tests on malicious applications and, at the same time, record user interaction
with the UI of the application. Our second goal is to automatically exercise
unknown applications, leveraging \straces{} previously recorded on similar
applications.

Our approach is to let applications run on a remote sandbox while users
seamlessly interact with their UI as if they were running locally on their
devices. More precisely, in \textbf{Phase~1 (Recording of \straces)}, each
sandbox uses a remote framebuffer protocol to collect \textit{\straces{}}, which
represent the sequence of UI events performed by the user, as well as the list
of UI elements actually stimulated during the test. Differently from previous
work (e.g., \cite{reran}), we go beyond recording raw events from
\texttt{/dev/input} and re-injecting them to another input device: We correlate
such events to the respective UI elements (e.g., buttons, or other view
objects), and collect information about the behaviors exhibited by the exercised
applications, through dynamic analysis. As described in
Section~\ref{sec:system-details}, this entails some challenges that we need to
solve. From hereinafter, a \emph{behavior} is a sequence of
observable runtime events (e.g., system calls, API calls).

\begin{figure}[t]
  \centering
  \includegraphics[width=\columnwidth]{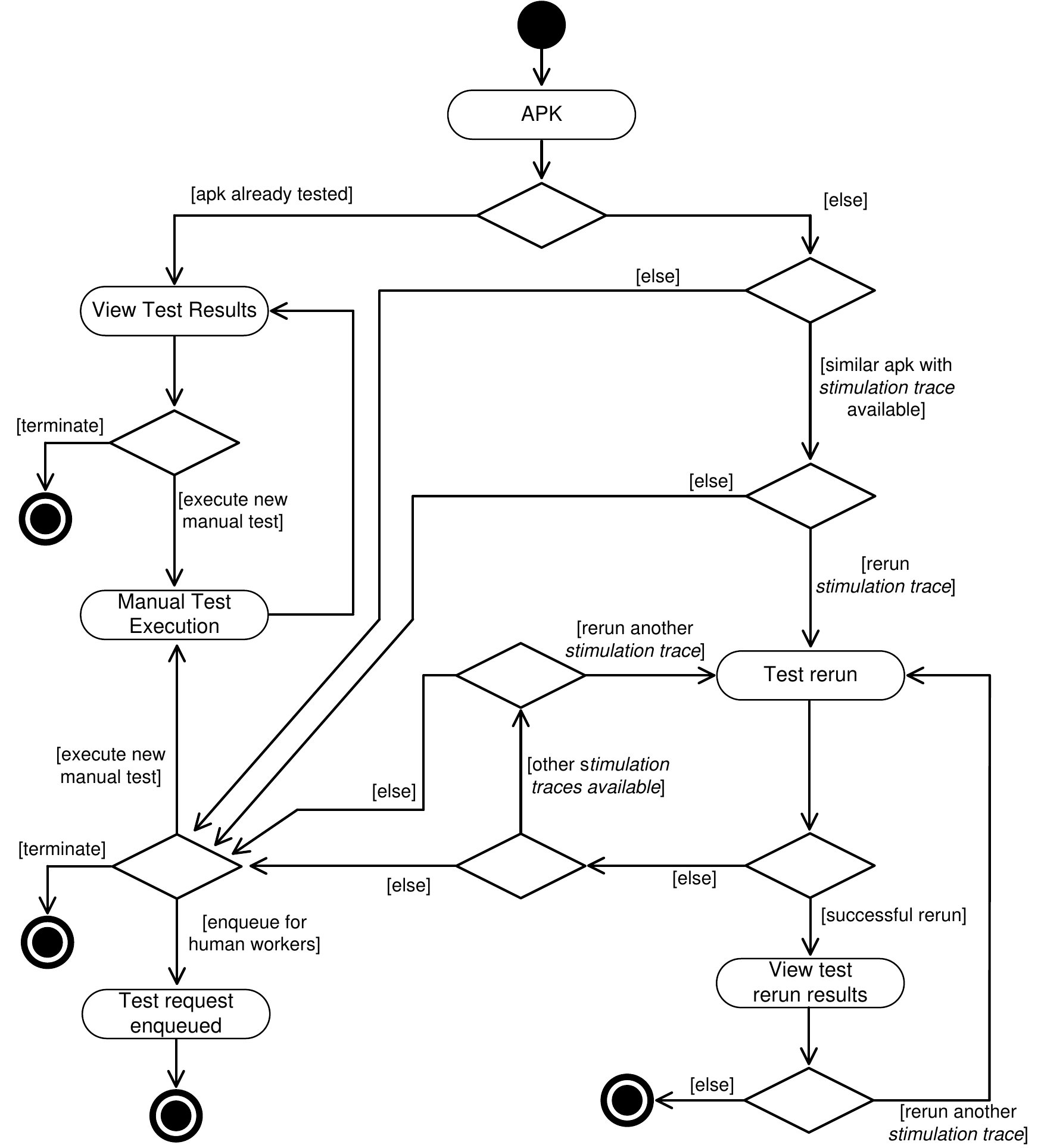}
  \caption{Workflow of our approach (Section~\ref{sec:system-details}).}
  \label{fig:workflow}
\end{figure}

Manual stimulation on large datasets is clearly unfeasible. Thus, in
\textbf{Phase~2 (Re-execution of \straces)}, we leverage the collected
\straces{} to automatically exercise new applications, so as to increase the
code covered during dynamic analysis. Our hypothesis is that by re-using
\straces{} we obtain better results (in terms of discovered behaviors) than with
random UI exercisers. A na\"ive approach where we blindly try to exercise an
application with \textit{every} \strace{} in our system is not accurate or
efficient. Therefore, in \textbf{Phase~3 (Finding Similar Applications)}, we
leverage the concept of \textit{UI application similarity}. As summarized by the
workflow in Figure~\ref{fig:workflow}, when a new sample is to be analyzed, we
first look for similar (or equivalent) samples for which we have a
\strace{}. Then, we use only \straces{} of the most similar known
application. With our approach, calculating the similarity between two
applications takes constant time and memory, whereas finding the most similar
application to a given one, in a database of $N$ applications, takes logarithmic
time.

\medskip\noindent One could argue that the need of collecting a set of
\straces{} large enough to be useful may limit the scalability of our
approach. However, we consider two factors. First, our system could attract the
interest not only of security analysts, but also of normal users that want to
safely try potentially malicious applications they find on the web or in
alternative markets. Secondly, we can leverage the accessibility of
crowdsourcing services, like Amazon Mechanical Turk, to recruit human workers and
generate new \straces{} efficiently (being this a matter of engineering and
deployment, we focus on our novel approach)

\section{System Details}
\label{sec:system-details}
\subsection{Phase~1: Recording of \straces}
\label{sec:input_recording}
We record the low-level input events generated while the user interacts with an
application on his or her device. We developed an extended VNC client and server
architecture through which this process happens transparently, with no changes
in the way users interact with the UI of an application. For the client, we
extended TightVNC, whereas we implemented the server on top of the Fastdroid
libraries\footnote{\url{https://code.google.com/p/fastdroid-vnc/}}.

We translate the input events in a sequence of remote framebuffer
(RFB)\footnote{\url{https://tools.ietf.org/rfc/rfc6143.txt}} \emph{PointerEvent}
or \emph{KeyEvent} messages that are sent to the VNC server. For each event, we
save the \verb|timestamp| (according to the client), \verb|event_type| (0 for
touch events and 1 for keys), \verb|action| (0 is ``up'', 1 is ``down'', 2 is
``move''), \verb|x_pos,y_pos| coordinates on the screen, and \verb|key_code|
(pressed button). Figure~\ref{fig:input_ev_example} shows an excerpt of a sample
input events file generated by the VNC server.

Two similar applications may have some subtle UI differences that can make a
re-execution test fail (e.g., slightly shifted buttons). Taking for example two
distinct BaseBridge samples\footnote{MD5s:
  00c154b42fd483196d303618582420b89cedbf46,
  73bb65b2431fefd01e0ebe66582a40e74928e053} from the Malware Genome
Project~\cite{malgenome}, we notice that the main button of the second sample is
slightly shifted. Exercising the second sample with the sequence of raw input
events recorded on the first sample (as one would do by using, for example, the
approach in~\cite{reran}), would cause an error.

To solve this, during recording we keep track of which view object (e.g., button
identifier) consumed each input event during recording, in order to find that
same view object during re-execution. For this, we rely on the
\texttt{ViewServer}, which allows to ``walk'' the
hierarchy\footnote{\url{http://developer.android.com/guide/topics/ui}} of
displayed objects. More precisely, our VNC server performs the following steps
when a new input event is received:

\begin{figure}[t]
  \centering
  \includegraphics[width=.7\columnwidth]{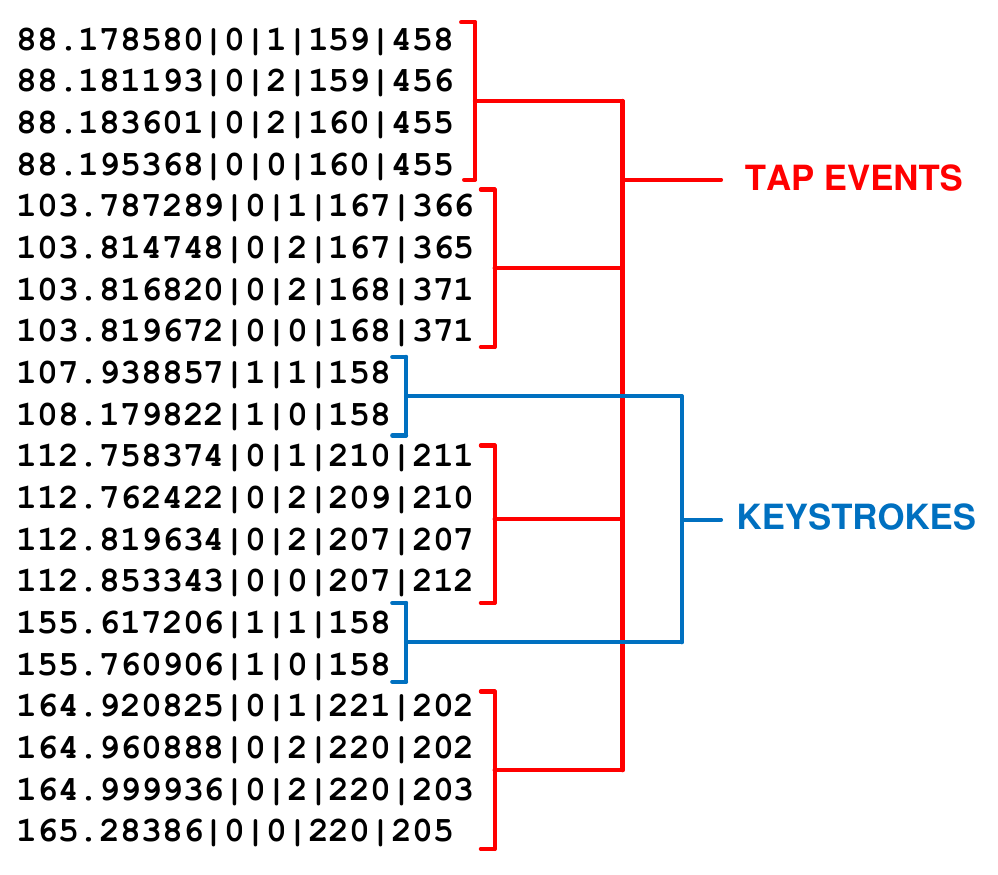}
  \caption{Excerpt of a sample input events file.}
  \label{fig:input_ev_example}
\end{figure}

\begin{figure}[b]
  \centering
  \includegraphics[width=.6\columnwidth]{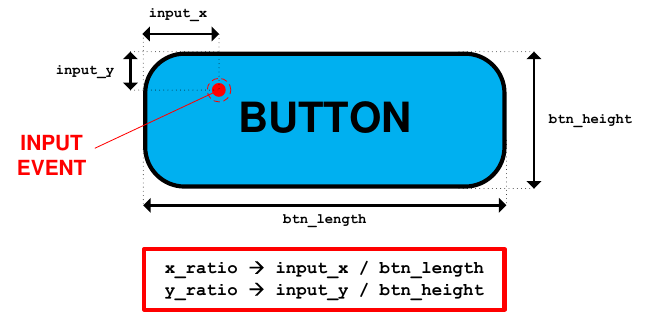}
  \caption{Input event relative position with respect to view object.}
  \label{fig:view_input_ratio}
\end{figure}

\begin{enumerate}\itemsep0em
\item Process RFB \emph{PointerEvent} message.
\item Send the \verb|GET_FOCUS| command to the \texttt{ViewServer}, to get the
  name and hash code of the focused window (i.e., \texttt{Activity}).
\item Retrieve view hierarchy of the window sending \texttt{DUMPQ} command to
  \texttt{ViewServer}.
\item Search view hierarchy for the deepest-rightmost view object
  containing the coordinates of the input event.
\item Store the paths to the previously found view nodes.
\end{enumerate}
We combine the collected information to extract the list of paths to the views
that actually consumed the touch events. Moreover, as motivated in
\textbf{Phase~2}, in case of touch events we log which activity has consumed
each event, and the path to all the deepest nodes in the hierarchy that can
consume the touch event. By combining this information with the coordinates of
the touch events generated by the user we build the sequence of view objects
stimulated. We also retain the relative position of the input event with respect
to the view object, which is useful in \textbf{Phase~2}.

\subsection{Phase~2: Re-execution of \straces}
\label{sec:rerunner}
For each event in the recorded sequence, we use the view object and ratio
information (see Figure~\ref{fig:view_input_ratio}) to properly re-scale the
horizontal and vertical coordinates. Then, we write the resulting event into the
\texttt{/dev/input} device.

\begin{figure}[t]
  \centering
  \includegraphics[width=\columnwidth]{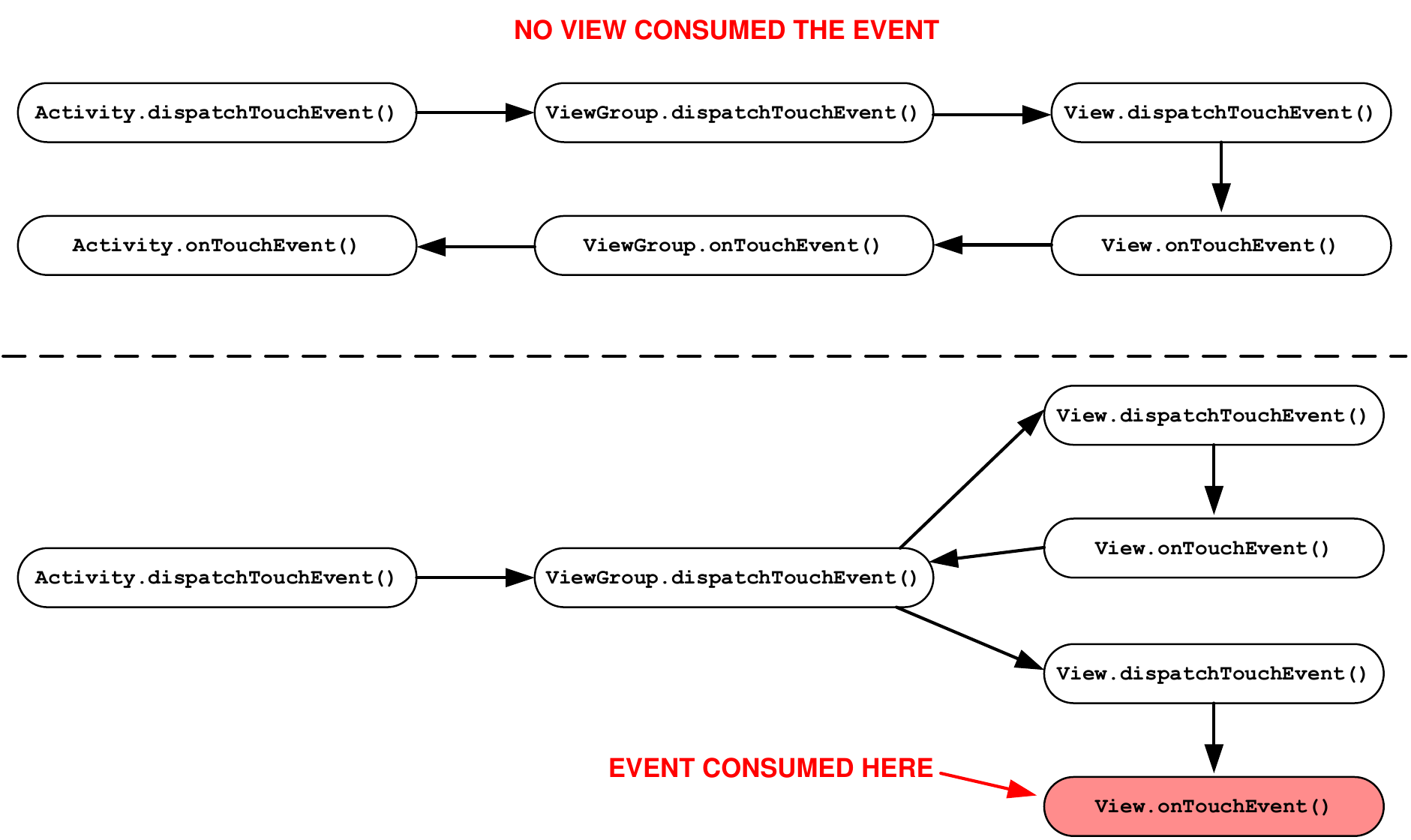}
  \caption{Examples of touch event management.}
  \label{fig:event_consuming}
\end{figure}

We treat touch events with special care to avoid the following rare corner
case, which can occur if the view that receives the input event is not the
view that eventually consumes it. More precisely, when a touch event is handled
by the Android Touch System~\cite{android_touch_sys}, the
\texttt{Activity.\-dispatch\-Touch\-Event()} method of the currently running
Activity is called. This method dispatches the event to the root view in the
hierarchy and waits for the result: If no view consumes the event, the Activity
calls \texttt{onTouchEvent()} in order to consume itself the event before
terminating. When a view object receives a touch event, the
\texttt{View.\-dispatch\-Touch\-Event()} is called: This method first tries to
find an attached listener to consume the event, calling
\texttt{View.\-On\-Touch\-Listener.\-onTouch()}, then tries to consume the event
itself calling \texttt{View.\-on\-Touch\-Event()}. If neither there is a
listener nor the \texttt{onTouchEvent()} method is implemented, the event is not
consumed and it flows back to the parent. When a \texttt{ViewGroup} receives a
touch event, it iterates on its children views in reverse order and, if the
touch event is inside the view, it dispatches the event to the child. If the
event is not consumed by the child, it continues to iterate on its children
until a view consumes the event. If the event is not consumed by any of its
children, the \texttt{ViewGroup} acts as a \texttt{View} and tries to consume
itself the event. Eventually, if it is not able to consume the event it sends
back to the parent. Figure~\ref{fig:event_consuming} shows two examples of
touch events management: In the former, the event flows down through the
hierarchy, and since it is not consumed by any view, it goes back to the
\texttt{Activity}. In the latter, the event is consumed by the second
\texttt{View} child of the \texttt{ViewGroup} object.

Our system avoids this corner case because it recorded, during \textbf{Phase~2},
which activity has consumed each event, and the path to all the deepest nodes in
the hierarchy that can consume the touch event.

\subsection{Phase~3: Finding Similar Applications}
\label{sec:similarity}
In case a \strace for an application $A$ is not available, after searching
by MD5,
we rely on visual similarity to find similar applications.

The goal of this phase is to find an application, $B$, for which a \strace
exists, and that has a UI similar to $A$s. To this end we leverage the concept
of visual similarity, implemented through perceptual hashing. Given an image in
input, a perceptual hashing algorithm creates a metric fingerprint that is
robust to image re-scaling, rotation, deformation, skew and compression. Thus,
if two images are visually similar, their respective hashes, which are 64-bits
unsigned integers, are very close. In particular, perceptually similar images
have a hamming distance within bounds that can be reliably
estimated\footnote{\url{http://phash.org/docs/design.html}}, as we also show in
Section~\ref{sec:ui-similarity-evaluation}.

In practice, to lookup a suitable \strace for application $A$, we calculate the
perceptual hash of its screenshots. Then, we look for $B$, an application which
screenshots minimize the hamming distance from $A$'s screenshots according to
their respective perceptual hash. We pre-calculate the hashes of the known
applications offline (which takes only 5.030453ms on average), and index them in
a MVP tree~\cite{Bozkaya02indexinglarge}, which allows lookup in logarithmic
time.

If a screenshot is already available, which is very likely if the application is
obtained from a market (e.g., screenshots are part of the app's metadata), our
our system calculates the perceptual hash using the \verb|ph_dct_imagehash|
function of the \texttt{libphash} library. In case no screenshot is available,
our system instantiates an emulator, installs the APK of $A$ and leverages the
\texttt{screencap} utility to take a screenshot once the application has
started.

\subsection{Implementation Details}
\label{sec:implementation}
The actual execution of the target application happens on the \emph{server
  tier}, which receives the UI events, and records and proxies them to an
instrumented Android Virtual Device (AVD) with the same screen size of the
client. AVDs are concurrently instantiated for each new client. A VNC server
instance is connected to each AVD screen to record the \straces. For
\textbf{Phase~2} the life cycle is almost the same, with the only differences
that, instead of connecting VNC server, we inject the re-scaled and adapted
input events into the running AVD. The devices associated to the touchscreen and
to the keyboard are fixed and respectively are \texttt{/dev/input/event1} and
\texttt{/dev/input/event2}.

We have tested \thesystem{} with the original AVD and DroidBox~\cite{droidbox}.
For our experiments we obtained access to
\copperdroid\footnote{\url{http://copperdroid.isg.rhul.ac.uk}}~\cite{copperdroid},
which allows automatic dynamic- and stimulation-based behavioral analysis.

We patched the \texttt{ViewServer}~\cite{patches} and stripped it down so as to
collect only  data useful to our purposes. This resulted in a 20--40x
speedup over the original implementation.


%% file: evaluation.tex
\section{Experimental evaluation}
Our results shows that both manual exercising and re-execution of collected
\straces reach higher code coverage than the one obtained with automatic UI
exercisers: We succeeded in stimulating more than the amount of behaviors
stimulated by other exercising strategies. Moreover, we found some particular
cases in which \thesystem{} succeeds in stimulating interesting malicious
behaviors that are not exposed using automatic application exercising
approaches.

In \textbf{Experiment~1} we verified that our stimulation approach led to a
better stimulation compared to other automatic analysis approaches. For this, we
compared the number of behaviors exercised with \thesystem vs. the ones
exercised with automatic approaches (i.e., monkey) typically used in dynamic
malware analysis frameworks, and vs. the system events stimulation strategy
proposed in \copperdroid~\cite{copperdroid}. In \textbf{Experiment~2} we
verified that the same \strace can be reused on similar samples to exercise the
same behaviors. For this, we compared the behaviors exercised on
manually-stimulated APKs with the behaviors exercised on similar samples, and
verified the outcome manually. In \textbf{Experiment~3} we verified that our UI
similarity approach is accurate and efficient.

For dynamic analysis, we obtained access to the \copperdroid sandbox, which is
convenient for our needs because (1) works at system-call, (2) incorporates a
state-of-the-art stimulation approach, able to stimulate both statically and
dynamically registered broadcast receivers, and (3) already provides a list of
behaviors built by means of system calls.

\subsection{Experiment~1: Manual UI exercising}
\label{sec:human}

\subsubsection{Dataset}
\label{sec:stim_eval_dataset}
We used 15 APKs samples, 13 from the Android Malware Genome
Project~\cite{malgenome}, and 2 from the Google Play store. The dataset is
purposely small, because we performed multiple tests on each sample and manually
inspected the output of each test in order to examine precisely the differences
between different approaches. Therefore, we preferred focusing on a
small dataset to perform a deeper analysis of each test result.

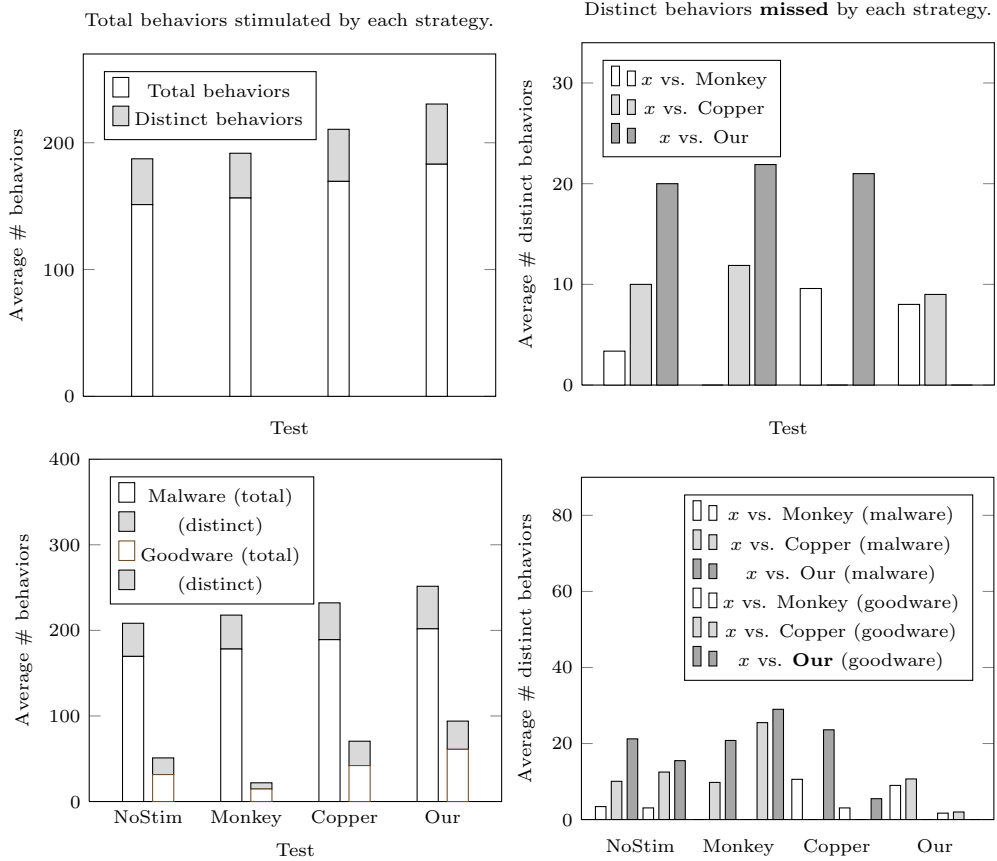
\begin{figure*}[t]
  \centering
  \begin{tikzpicture}
    \begin{axis}[%
      scale=0.8,
      legend style={at={(0.05,0.95)}, anchor=north west, legend columns=1},
      symbolic x coords={NoStim, Monkey, Copper, Our},
      xticklabel=\empty,
      xtick style={draw=none},
      title={Total behaviors stimulated by each strategy.},
      ylabel={Average \# behaviors},
      xlabel={Test},
      enlarge x limits=0.2,
      ybar stacked,
      ymin=0,
      ymax=270,
      bar width=8pt,]

      \addplot +[black, fill=white]
      coordinates{
        (NoStim, 151.2)
        (Monkey, 156.39333333346667)
        (Copper, 169.53333333333333)
        (Our, 183.06666666666666) };

      \addplot +[black, fill=gray!30!]
      coordinates{
        (NoStim,36.13333333333333)
        (Monkey,35.31666666666667)
        (Copper,41.06666666666667)
        (Our,47.46666666666667) };

      \legend{Total behaviors, Distinct behaviors}
    \end{axis}
  \end{tikzpicture}
  \begin{tikzpicture}
    \begin{axis}[%
      scale=0.8,
      legend style={at={(0.05,0.95)}, anchor=north west,legend columns=1},
      symbolic x coords={NoStim, Monkey, Copper, Our},
      xticklabel=\empty,
      xtick style={draw=none},
      ylabel={Average \# distinct behaviors},
      xlabel={Test},
      title={Distinct behaviors \textbf{missed} by each strategy.},
      enlarge x limits=0.2,
      ybar,
      ymax=34,
      ymin=0,
      bar width=8pt]

      \addplot [black,fill=white]
      coordinates{
        (NoStim, 3.37)
        (Monkey, 0)
        (Copper, 9.59)
        (Our, 8.01) };

      \addplot [black,fill=gray!30!]
      coordinates{
        (NoStim, 10)
        (Monkey, 11.88)
        (Copper, 0)
        (Our, 9) };

      \addplot [black,fill=gray!70!]
      coordinates{
        (NoStim, 20)
        (Monkey, 21.9)
        (Copper, 21)
        (Our, 0) };

      \legend{$x$ vs. Monkey, $x$ vs. Copper, $x$ vs. Our}
    \end{axis}
  \end{tikzpicture}

  \begin{tikzpicture}
    \begin{axis}[
      scale=0.8,
      legend style={at={(0.05,0.95)}, anchor=north west, legend columns=1},
      symbolic x coords={NoStim, Monkey, Copper, Our},
      xtick=data,
      xtick style={draw=none},
      ylabel={Average \# behaviors},
      xlabel={Test},
      enlarge x limits=0.2,
      ybar stacked,
      ymin=0,
      ymax=400,
      bar width=8pt,]

      \addplot +[black,fill=white,bar shift=-.2cm]
      coordinates{
        (NoStim,169.615384615)
        (Monkey,178.203846154)
        (Copper,189.153846154)
        (Our,201.846153846) };

      \addplot +[black,fill=gray!30!,bar shift=-.2cm]
      coordinates{
        (NoStim,38.6923076923)
        (Monkey,39.6307692308)
        (Copper,43.0)
        (Our,49.6923076923) };

      \resetstackedplots

      \addplot +[fill=white,postaction={pattern=north east lines},bar shift=.2cm]
      coordinates{
        (NoStim,31.5)
        (Monkey,14.625)
        (Copper,42.0)
        (Our,61.0) };

      \addplot +[fill=gray!30!,postaction={pattern=north east lines},bar shift=.2cm]
      coordinates{
        (NoStim,19.5)
        (Monkey,7.275)
        (Copper,28.5)
        (Our,33.0) };

      \legend{Malware (total), (distinct), Goodware (total), (distinct)}
    \end{axis}
  \end{tikzpicture}
  \begin{tikzpicture}
    \begin{axis}[%
      scale=0.8,
      x tick label style={/pgf/number format/1000 sep=},
      legend style={at={(0.95,0.95)}, anchor=north east,legend columns=1},
      symbolic x coords={NoStim, Monkey, Copper, Our},
      xtick=data,
      xtick style={draw=none},
      enlarge x limits=0.2,
      ylabel={Average \# distinct behaviors},
      ybar,
      ymin=0,
      ymax=90,
      bar width=4pt,]

      \addplot [black,fill=white]
      coordinates{(NoStim,3.42) (Monkey,0) (Copper,10.59) (Our,8.99) };

      \addplot [black,fill=gray!30!]
      coordinates{(NoStim,10.07) (Monkey,9.78) (Copper,0) (Our,10.69) };

      \addplot [black,fill=gray!70!]
      coordinates{(NoStim,21.23) (Monkey,20.81) (Copper,23.61) (Our,0) };

      \addplot [black,fill=white,postaction={pattern=north east lines}]
      coordinates{(NoStim,3.075) (Monkey,0) (Copper,3.075) (Our,1.7) };

      \addplot [fill=gray!30!,postaction={pattern=north east lines}]
      coordinates{(NoStim,12.5) (Monkey,25.5) (Copper,0) (Our,2) };

      \addplot [fill=gray!70!,postaction={pattern=north east lines}]
      coordinates{(NoStim,15.5) (Monkey,29) (Copper,5.5) (Our,0) };

      \legend{%
        $x$ vs. Monkey (malware),
        $x$ vs. Copper (malware),
        $x$ vs. Our (malware),
        $x$ vs. Monkey (goodware),
        $x$ vs. Copper (goodware),
        $x$ vs. \textbf{Our} (goodware),}
    \end{axis}
  \end{tikzpicture}
  \caption{(left) Total behaviors stimulated on average by each strategy, and
    (right) behavior missed by each strategy. (bottom) breakdown of the
    above experiment on malware (m/w) and goodware (g/w).}
  \label{fig:experiment1}
\end{figure*}

\subsubsection{Experimental setup and procedure}
\label{sec:stim_eval_setup}
For each sample in our dataset, we collect the system call traces during
execution with four stimulation approaches:

\begin{itemize}\itemsep0em
\item \textbf{NoStim}: without stimulation,
\item \textbf{Copper}: with \copperdroid{} stimulation strategy,
\item \textbf{Monkey} stimulation with an increasing number of input events (500,
  1000, 2000, 5000).
\item \textbf{Our}: an everyday Android user exercised the sample through
  \thesystem{}, without knowing the outcome of the other tests. We
  instructed the user to rely on his sole knowledge and try to use the
  application naturally, following anything  the application asks,
  without thinking whether the action is dangerous or not.
\end{itemize}
For each couple of stimulation approaches A and B, we calculate the total
stimulated behaviors by A and B, and the behaviors stimulated only by either A
or B (set difference). In each test, we start a clean sandbox, install the APK
sample, run it with the selected stimulation approach, and collect the system
calls traces and the behavior lists.

\subsubsection{Results}
\label{sec:stim_eval_results}
We calculated the average number of behaviors (total and distinct) observed with
each stimulation approach, and the average number of distinct missed behaviors
by each strategy (set difference). We repeated the experiment on the entire
dataset, and then on each set of goodware and malware applications.

Figure~\ref{fig:experiment1} summarizes the results. From the bar
chart on the top-left corner, we can see that the human-driven stimulation
succeeds in stimulating more behaviors than any automatic approaches: we are
able to exercise 112\% of total behaviors and 124\% of distinct behaviors more
than the automatic stimulation. The same result holds regardless of whether the
application is malicious or benign. The bar chart on the bottom-left corner
confirms the above results regardless of whether the applications are benign
(striped bars) or malicious (solid bars).

From the bar chart on the top-right corner, we can see that the other approaches
miss many behaviors with respect to ours, whereas our technique misses a
negligible amount of behaviors. From another perspective, \thesystem{} is able
to stimulate 593\% exclusive behaviors more in respect to monkey and 200\% more
i respect to the state of the art (\copperdroid). The bar chart on the
bottom-right corner confirms the above results regardless of whether the
applications are benign (striped bars) or malicious (solid bars). We analyze the
results on malware and goodware samples separately.

Overall, our results confirm our hypothesis that \thesystem{} UI stimulation
approach allows to obtain better results than automatic approaches during
dynamic analysis.

\begin{figure}[t]
  \centering
  \includegraphics[width=.9\columnwidth]{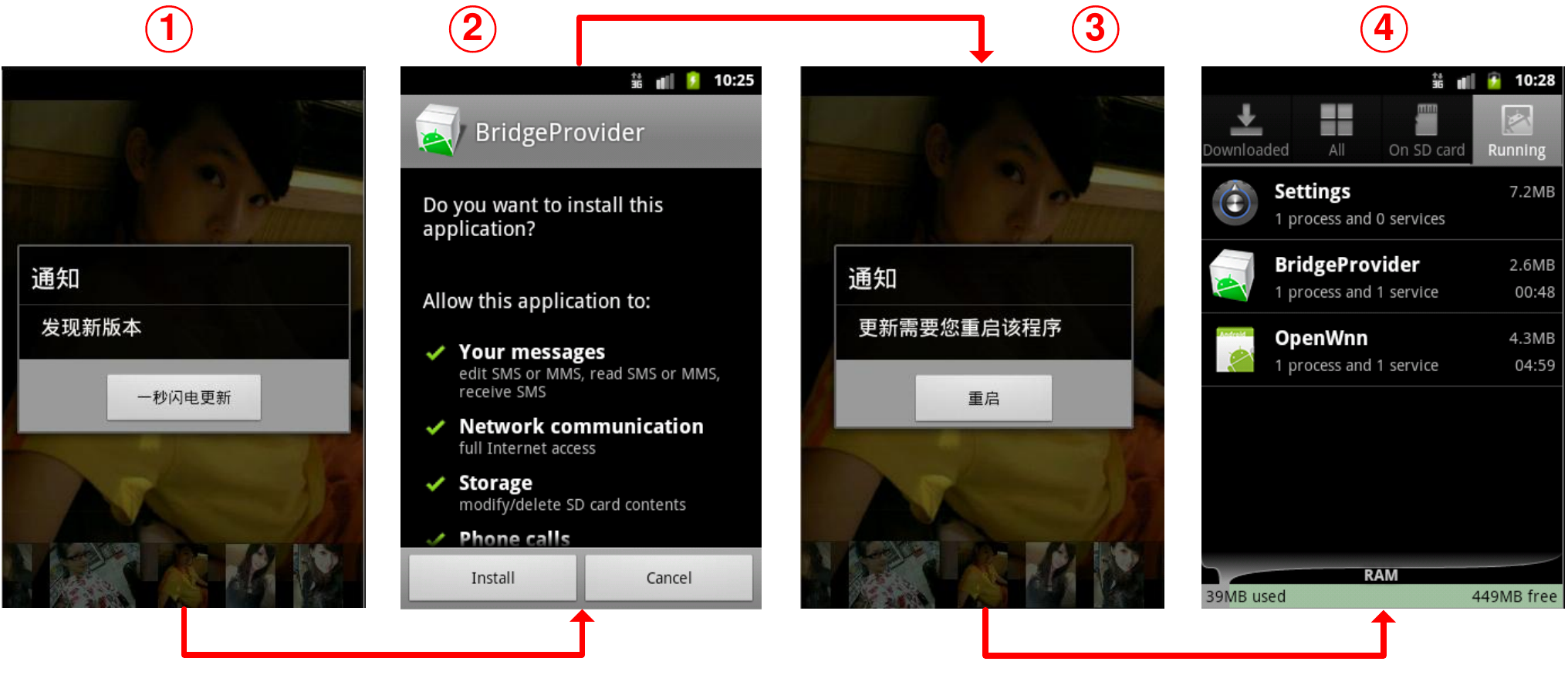}
  \caption{Experiment~1 (Case analysis): Steps to install BridgeProvider
    payloads: 1) Ask for application update; 2) Install payload; 3) Restart
    application; 4) Malicious service running on device.}
  \label{fig:basebridge_steps}
\end{figure}

\paragraph*{\textbf{Case analysis}}
A notable case that deserves detailed analysis is that of a malicious behavior
exercised, and thus exposed, exclusively by our system. The malware sample under
analysis is \texttt{com.keji.danti80}, belonging to BaseBridge malware
family. BaseBridge is a trojan that, once installed, prompts the user with an
upgrade dialog: if users accept to do so, the malware will install a second
malicious application on the phone. This service communicates with a control
server to receive instructions to perform unwanted activities (e.g., place calls
or send messages to premium numbers). Meanwhile, the malware also blocks
messages from the mobile carrier in order to prevent users from getting fee
consumption updates: in this way all malicious activities are undertaken
stealthily without the users' knowledge or consent. More details on this malware
can be found in~\cite{basebridge_mobmal, basebridge_mcafee,
  basebridge_symantec}. This case is very common and is used by malware authors
to circumvent dynamic analysis with ineffective UI exercising.

Analyzing the sample with \thesystem{} we obtained the list of behaviors shown
in Table~\ref{tbl:stim_eval_case}. The underlined lines indicate a behavior
that none of the other stimulation techniques were able to reveal. The malware
writes another APK file, \texttt{xxx.apk}, on the filesystem. As a matter of
fact, during the test, the application prompts the user to install a new
malicious application, named BridgeProvider, to complete the update, as shown in
Figure~\ref{fig:basebridge_steps}.

\begin{table}[t]
  \scriptsize
  \centering
  \caption{Experiment~1: List of behaviors extracted testing
    \texttt{com.keji.danti80} malware sample.}
  \setlength{\tabcolsep}{1.5pt}
  \sffamily
  \renewcommand{\arraystretch}{0.65}
  \begin{tabular}{crl}
    \toprule
    \textbf{Hits} & \textbf{Beh.} & \textbf{Data} \\
    \midrule
    2 & write & {\scriptsize /sys/qemu\_trace/process\_name}\\
    2 & write & {\scriptsize /sys/qemu\_trace/process\_name}\\
    2 & write & {\scriptsize /sys/qemu\_trace/process\_name}\\
    1 & \underline{mkdir} & \underline{{\scriptsize /data/com.keji.danti80/files}}\\
    1 & \underline{write} & \underline{{\scriptsize /data/com.keji.danti80/files/xxx.apk}}\\
    2 & write & {\scriptsize /sys/qemu\_trace/process\_name}\\
    2 & write & {\scriptsize /sys/qemu\_trace/process\_name}\\
    1 & \underline{mkdir} & \underline{{\scriptsize /data/com.sec.android.bridge/shared\_prefs}}\\
    1 & \underline{unlink} & \underline{{\scriptsize /data/com.keji.danti80/files/xxx.apk}}\\
    1 & connect & {\scriptsize host: 10.0.2.3, retval: 0, port: 53}\\
    1 & ns\_query & {\scriptsize query\_data: b3.8866.org. 1 1}\\
    1 & connect & {\scriptsize host: 221.5.133.18, retval: -115, port: 8080}\\
    1 & write & {\scriptsize /data/com.sec.android.bridge/shared\_prefs/first\_app\_perfs.xml}\\
    3 & unlink & {\scriptsize /data/com.sec.android.bridge/shared\_prefs/first\_app\_perfs.xml.bak}\\
    1 & write & {\scriptsize /data/com.keji.danti80/files/atemp.jpg}\\
    1 & unlink & {\scriptsize /data/com.keji.danti80/files/atemp.jpg}\\
    2 & write & {\scriptsize 221.5.133.18 port:8080}\\
    \bottomrule
  \end{tabular}
  \label{tbl:stim_eval_case}
\end{table}

In conclusions, other stimulation approaches did not exercise the malware enough
to make it reveal its true malicious behavior, with the consequent risk to
consider the sample as safe. Instead, using \thesystem{}, the analyst is able to
detect such a potential dangerous behavior and subsequently analyze in detail
the functioning of the application.

\medskip\noindent\textbf{Conclusions of Experiment~1:} The results confirmed our
intuition that automatic UI stimulation approaches can only exercise a subset of
the (malicious) behaviors of a malware during dynamic analysis.
Moreover, \thesystem{} approach based on human-driven UI exercising allows to
reproduce typical victim interaction with the malware and to reach then higher
code coverage.

\subsection{Experiment~2: Automatic Re-execution}
\label{sec:scalability_eval}
This experiment's goal is to verify the following novel hypothesis: If we
succeed in exercising the (malicious) behaviors in a sample, the same
stimulation should trigger behaviors in a similar sample.

In a preliminary experiment, we measured that the percentage of UI events
successfully re-executed on a dataset of similar applications is 88.52\%. This
percentage is actually a conservative estimate. For example, suppose that we
have a recording of a UI stimulation with 20 events: if \thesystem succeeds in
re-executing 10 UI events but it is not able to find the correct view to inject
the 11\textsuperscript{th} event, the re-execution is terminated. We then have a
re-execution score of 50\%.

In the reminder of this section we show the impact of such re-execution on
the behaviors exposed during dynamic analysis.

\begin{table}[t]
  \scriptsize
  \centering
  \caption{Experiment~2: Summary of the results (average values per test).}
  \begin{tabular}{lcccc}
    \toprule
    \multicolumn{5}{c}{\textbf{\emph{a) ManualVsRe-exec}}}  \\
    \midrule
    \multicolumn{2}{l}{ \textbf{Manual test}} & \multicolumn{3}{c}{201.85 (38.69 distinct)} \\
    \multicolumn{2}{l}{ \textbf{Re-executed tests}} & \multicolumn{3}{c}{230.20 (52.76 distinct)} \\
    \multicolumn{2}{l}{ \textbf{Only in manual}} & \multicolumn{3}{c}{24.00} \\
    \multicolumn{2}{l}{ \textbf{Only in re-executed}} & \multicolumn{3}{c}{25.00} \\

    \midrule

    \multicolumn{5}{c}{\textbf{\emph{c) AutomaticVsRe-exec}}}  \\
    \toprule
    \multicolumn{5}{c}{\textbf{Stimulated Behaviors}}  \\
    \midrule
    \multicolumn{2}{l}{\textbf{No Stimulation}} & \multicolumn{3}{c}{199.79 (39.73 distinct)}\\
    \multicolumn{2}{l}{\textbf{Monkey}} & \multicolumn{3}{c}{196.62 (39.61 distinct)}\\
    \multicolumn{2}{l}{\textbf{CopperDroid}} & \multicolumn{3}{c}{198.95 (41.84 distinct)}\\
    \multicolumn{2}{l}{\textbf{Our}} & \multicolumn{3}{c}{230.20 (52.76 distinct)}\\

    \midrule

    \multicolumn{5}{c}{ \textbf{Exclusive Behaviors}}  \\
    \toprule
     \textbf{Only in} & { \textbf{NoStim}} & { \textbf{Monkey}} & { \textbf{Copper}} & { \textbf{Our}} \\
    \midrule
     \textbf{Monkey} & 4.35 & 0 & 8.37 & 8.82 \\
     \textbf{Copper} & 6.54 & 8.05 & 0 & 7.74 \\
     \textbf{Our} & 23.27 & 22.58 & 22.15 & 0 \\
    \bottomrule
  \end{tabular}
  \label{tbl:scal_eval_sumup}
\end{table}

\subsubsection{Dataset}
\label{sec:scalability_eval_dataset}
For this experimental evaluation we picked 13 malware samples from the Android
Malware Genome Project~\cite{malgenome} used in \textbf{Experiment 1}, run
\textbf{Phase~1} to record \straces, and \textbf{Phase~3} to retrieve similar
APKs from a repository of over 7,000 samples that also include Google Play and
alternative markets.

\subsubsection{Experimental setup and procedure}
\label{sec:scalability_eval_setup}
To verify if our re-execution approach is feasible, we need a way to evaluate
the results of the re-executed tests. We follow four criteria:
\begin{itemize}\itemsep0em
\item [a)] \emph{\textbf{ManualVsRe-exec:}} Compare the behaviors exercised with
  manual stimulation vs. the behaviors exercised automatically.
\item [b)] \emph{\textbf{Re-execBehaviors:}} Verify if an interesting malicious
  behavior exhibited in the original sample is also exhibited during the
  re-execution on a similar application.
\item [c)] \emph{\textbf{AutomaticVsRe-exec:}} Compare the behaviors exercised
  using automatic stimulation tools, as in \textbf{Experiment~1}, against the
  behaviors extracted during execution.
\end{itemize}
We structured each test as follows, for each of the 13 APK:
\begin{enumerate}\itemsep0em
\item As in \textbf{Experiment~1}, we ask a user to manually test the
  application while \thesystem records UI stimulation during traces.
\item Search for the most similar APK using \textbf{Phase~3}.
\item Run \textbf{Phase~2} to automatically re-execute previously recorded UI
  stimulation on the similar application.
\item Test the similar application with automatic stimulation approaches:
  \begin{itemize}
  \item \textbf{NoStim:} 1 test without stimulation,
  \item \textbf{Monkey:} 20 tests using monkey,
  \item \textbf{Copper:} 1 test using \copperdroid{} stimulation strategy.
  \end{itemize}
\item Calculate the four evaluation criteria explained above.
\end{enumerate}
For each sample in the dataset we performed one test.

\begin{figure*}[t]
  \centering
  \begin{tikzpicture}[scale=0.7]
    \tikzstyle{every node}=[font=\scriptsize]
    \begin{axis}[
      x tick label style={/pgf/number format/1000 sep=},
      nodes near coords=\rotatebox{90}{\pgfplotspointmeta},
      legend style={at={(0.95,0.95)}, anchor=north east, legend columns=-1},
      symbolic x coords={1, 2, 3, 4, 5, 6, 7, 8, 9, 10, 11, 12, 13},
      xtick=data,
      enlarge x limits=0.2,
      ybar,
      ymax=2000,
      ymin=0,
      ylabel={Average total \# behaviors},
      title={Total behaviors (average)},
      bar width=4pt,]

      \addplot +[black, fill=gray!30!, point meta=explicit symbolic]
      coordinates{
        (1,882.00) [882]
        (2,126.00) [126]
        (3,64.00) [64]
        (4,161.00) [161]
        (5,68.00) [68]
        (6,73.00) [73]
        (7,55.00) [55]
        (8,59.00) [59]
        (9,746.00) [746]
        (10,97.00) [97]
        (11,124.00) [124]
        (12,92.00) [92]
        (13,158.00) [158]};

      \addplot +[black, fill=gray!60!, point meta=explicit symbolic]
      coordinates{
        (1,1250.00) [1250.00 (141\%)]
        (2,102.00) [102.00 (80\%)]
        (3,81.00) [81.00 (126\%)]
        (4,132.00) [132.00 (81\%)]
        (5,92.00) [92.00 (135\%)]
        (6,81.00) [81.00 (110\%)]
        (7,88.00) [88.00 (160\%)]
        (8,76.00) [76.00 (128\%)]
        (9,759.00) [759.00 (101\%)]
        (10,58.00) [58.00 (59\%)]
        (11,94.00) [94.00 (75\%)]
        (12,79.00) [79.00 (85\%)]
        (13,97.00) [97.00 (61\%)] };

      \legend{Manual, Re-execution}
    \end{axis}
  \end{tikzpicture}
  \begin{tikzpicture}[scale=0.7]
    \tikzstyle{every node}=[font=\scriptsize]
    \begin{axis}[
      x tick label style={/pgf/number format/1000 sep=},
      nodes near coords=\rotatebox{90}{\pgfplotspointmeta},
      legend style={at={(0.95,0.95)}, anchor=north east, legend columns=-1},
      symbolic x coords={1, 2, 3, 4, 5, 6, 7, 8, 9, 10, 11, 12, 13},
      xtick=data,
      enlarge x limits=0.2,
      ybar,
      ymax=360,
      ymin=0,
      title={Distinct behaviors (average)},
      ylabel={Average distinct \# behaviors},
      bar width=4pt,]

      \addplot +[black, fill=gray!30!, point meta=explicit symbolic]
      coordinates{
        (1,204.00) [204]
        (2,39.00) [39]
        (3,25.00) [25]
        (4,48.00) [48]
        (5,25.00) [25]
        (6,28.00) [28]
        (7,21.00) [21]
        (8,24.00) [24]
        (9,143.00) [143]
        (10,25.00) [25]
        (11,35.00) [35]
        (12,19.00) [19]
        (13,28.00) [28] };

      \addplot +[black, fill=gray!60!, point meta=explicit symbolic]
      coordinates{
        (1,221.00) [221.00 (108\%)]
        (2,37.00) [37.00 (94\%)]
        (3,27.00) [27.00 (108\%)]
        (4,45.00) [45.00 (93\%)]
        (5,28.00) [28.00 (112\%)]
        (6,26.00) [26.00 (92\%)]
        (7,25.00) [25.00 (119\%)]
        (8,26.00) [26.00 (108\%)]
        (9,146.00) [146.00 (102\%)]
        (10,22.00) [22.00 (88\%)]
        (11,30.00) [30.00 (85\%)]
        (12,19.00) [19.00 (100\%)]
        (13,29.00) [29.00 (103\%)] };

      \legend{Manual, Re-execution}
    \end{axis}
  \end{tikzpicture}
  \begin{tikzpicture}[scale=0.7]
    \tikzstyle{every node}=[font=\scriptsize]
    \begin{axis}[
      x tick label style={/pgf/number format/1000 sep=},
      nodes near coords=\rotatebox{90}{\pgfmathprintnumber\pgfplotspointmeta},
      legend style={at={(0.95,0.95)}, anchor=north east,legend columns=-1},
      symbolic x coords={1, 2, 3, 4, 5, 6, 7, 8, 9, 10, 11, 12, 13},
      xtick=data,
      enlarge x limits=0.2,
      ybar,
      ymax=240,
      ymin=0,
      title={Exclusive behaviors},
      bar width=4pt,]

      \addplot [black, fill=gray!30!]
      coordinates{
        (1,188.00)
        (2,1.00)
        (3,4.00)
        (4,13.00)
        (5,0.00)
        (6,2.00)
        (7,0.00)
        (8,1.00)
        (9,79.00)
        (10,2.00)
        (11,6.00)
        (12,11.00)
        (13,5.00) };

      \addplot [black, fill=gray!60!]
      coordinates{
        (1,206.00)
        (2,1.00)
        (3,6.00)
        (4,12.00)
        (5,3.00)
        (6,0.00)
        (7,2.00)
        (8,3.00)
        (9,78.00)
        (10,0.00)
        (11,2.00)
        (12,8.00)
        (13,6.00) };

      \legend{Manual, Re-execution}
    \end{axis}
  \end{tikzpicture}

  \caption{Experiment~2 (ManualVsRe-exec): Comparison of behaviors stimulated in
    the original, manual execution vs. the average total and distinct behaviors
    stimulated in re-executed tests. The third bar graph shows the exclusive
    behaviors exercised in either manual or re-executed tests.}
  \label{fig:scal_eval_comp_orig_rerun_1}
\end{figure*}
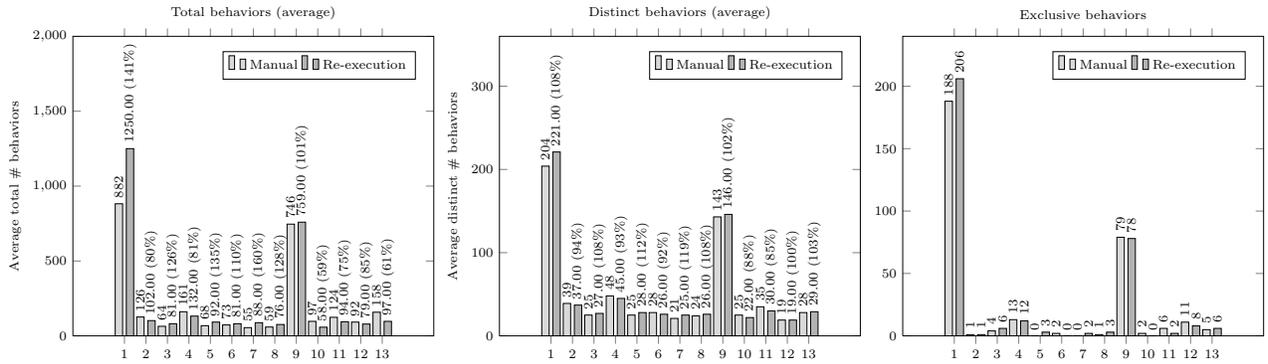

\begin{figure*}[t]
  \centering
  \begin{tikzpicture}[scale=0.7]
    \begin{axis}[
      x tick label style={/pgf/number format/1000 sep=},
      nodes near coords=\rotatebox{90}{\pgfmathprintnumber\pgfplotspointmeta},
      legend style={at={(0.5,-0.15)}, anchor=north,legend columns=-1},
      symbolic x coords={NoStim, Monkey, Copper, Puppet},
      xtick=data,
      enlarge x limits=0.2,
      ybar,
      ymax=250,
      ymin=0,
      title={Average distinct and total behaviors exercised},
      bar width=8pt,]
      \addplot [black,fill=white]
      coordinates{
        (NoStim,169.62)
        (Monkey,178.20)
        (Copper,189.15)
        (Puppet,201.85) };
      \addplot [black,fill=gray!60!]
      coordinates{
        (NoStim,38.69)
        (Monkey,39.63)
        (Copper,43.00)
        (Puppet,49.69)};

      \legend{Total, Distinct}
    \end{axis}
  \end{tikzpicture}
  \begin{tikzpicture}[scale=0.7]
    \begin{axis}[
      x tick label style={/pgf/number format/1000 sep=},
      nodes near coords=\rotatebox{90}{\pgfmathprintnumber\pgfplotspointmeta},
      legend style={at={(0.5,-0.15)}, anchor=north,legend columns=-1},
      symbolic x coords={NoStim, Monkey, Copper, Our},
      xtick=data,
      enlarge x limits=0.2,
      ybar,
      ymax=28,
      ymin=0,
      title={Distinct \textbf{missed} behaviors by each strategy},
      bar width=8pt,]
      \addplot [black,fill=white]
      coordinates{
        (NoStim,3.70)
        (Monkey,0)
        (Copper,5.88)
        (Our,8.42) };

      \addplot [black,fill=gray!30!]
      coordinates{
        (NoStim,6.54)
        (Monkey,8.05)
        (Copper,0)
        (Our,7.74) };

      \addplot [black,fill=gray!70!]
      coordinates{
        (NoStim,23.27)
        (Monkey,22.58)
        (Copper,22.15)
        (Our,0) };
      \legend{Monkey, Copper, Our}
    \end{axis}
  \end{tikzpicture}
  \caption{Experiment~2 (AutomaticVsRe-exec): Behaviors stimulated with
    re-execution in respect to behaviors extracted using automatic stimulation
    (left) and missed behaviors by each strategy (right).}
  \vspace*{5pt}
\label{fig:scal_eval_rerun_stim}
\end{figure*}
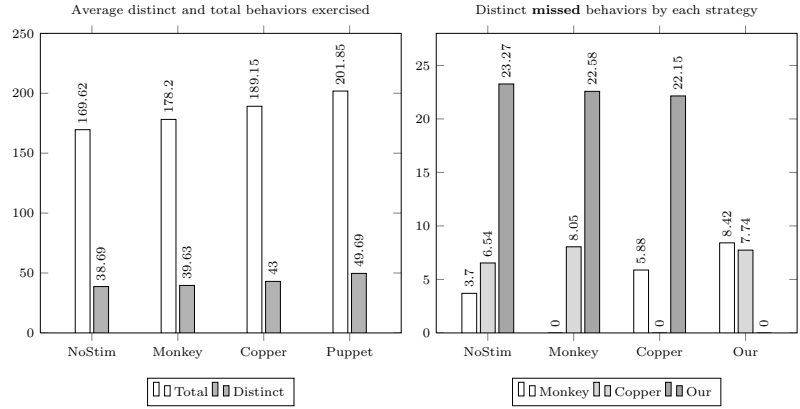

\begin{table*}[t!]
  \scriptsize
  \caption{Experiment~2 (Re-execBehaviors): Behaviors found
    on \texttt{com.keji.danti160} (left) and re-executed automatically
    from \thesystem on a similar sample (right).}
  \setlength{\tabcolsep}{1pt}
  \sffamily
  \renewcommand{\arraystretch}{0.65}
  \begin{tabular}{crlcrl}
    \toprule
    \multicolumn{3}{c}{Original execution} & \multicolumn{3}{c}{Re-execution on a similar sample}\\
    \midrule
    \textbf{Data} & \textbf{Beh.} & \textbf{Data} & \textbf{Hits} & \textbf{Beh.} & \textbf{Blob} \\
    \midrule
    2 & {write} & {\scriptsize /sys/qemu\_trace/process\_name} &                                        2 & {write} & {\scriptsize /sys/qemu\_trace/process\_name}\\
    2 & {write} & {\scriptsize /sys/qemu\_trace/process\_name} &                                        2 & {write} & {\scriptsize /sys/qemu\_trace/process\_name}\\
    2 & {write} & {\scriptsize /sys/qemu\_trace/process\_name} &                                        2 & {write} & {\scriptsize /sys/qemu\_trace/process\_name}\\
    1 & {mkdir} & {\scriptsize /data/com.keji.danti160/shared\_prefs} &                                 1 & {mkdir} & {\scriptsize /data/com.keji.danti161/shared\_prefs}\\
    1 & \underline{{mkdir}} & \underline{{\scriptsize /data/com.keji.danti160/files}} &                 1 & \underline{{mkdir}} & \underline{{\scriptsize /data/com.keji.danti161/files}}\\
    1 & \underline{{write}} & \underline{{\scriptsize /data/com.keji.danti160/files/xxx.apk}} &         1 & \underline{{write}} & \underline{{\scriptsize /data/com.keji.danti161/files/xxx.apk}}\\
    2 & {write} & {\scriptsize /sys/qemu\_trace/process\_name} &                                        2 & {write} & {\scriptsize /sys/qemu\_trace/process\_name}\\
    2 & {write} & {\scriptsize /sys/qemu\_trace/process\_name} &                                        2 & {write} & {\scriptsize /sys/qemu\_trace/process\_name}\\
    1 & \underline{{mkdir}} & \underline{{\scriptsize /data/com.sec.android.bridge/shared\_prefs}} &    2 & \underline{{mkdir}} & \underline{{\scriptsize /data/com.sec.android.bridge/shared\_prefs}}\\
    1 & {connect} & {\scriptsize host: 10.0.2.3, retval: 0, port: 53} &                                 2 & {write} & {\scriptsize /data/com.sec.android.bridge/shared\_ prefs/first\_app\_perfs.xml}\\
    1 & {ns\_query} & {\scriptsize query\_data: b3.8866.org. 1 1} &                                     1 & {unlink} & {\scriptsize /data/com.sec.android.bridge/shared\_ prefs/first\_app\_perfs.xml.bak}\\
    1 & \underline{{unlink}} & \underline{{\scriptsize /data/com.keji.danti160/files/xxx.apk}} &        1 & {connect} & {\scriptsize host: 10.0.2.3, retval: 0, port: 53}\\
    1 & {mkdir} & {\scriptsize /data/com.keji.danti160/databases} &                                     1 & {ns\_query} & {\scriptsize query\_data: b3.8866.org. 1 1}\\
    24 & {write} & {\scriptsize /data/com.keji.danti160/databases/db.db} &                              1 & \underline{{unlink}} & \underline{{\scriptsize /data/com.keji.danti161/files/xxx.apk}}\\
    3 & {write} & {\scriptsize /data/com.sec.android.bridge/shared\_prefs/first\_app\_perfs.xml} &      1 & {connect} & {\scriptsize host: 221.5.133.18, retval: -115, por t: 8080}\\
    2 & {unlink} & {\scriptsize /data/com.sec.android.bridge/shared\_prefs/first\_app\_perfs.xml.bak} & 1 & {mkdir} & {\scriptsize /data/com.keji.danti161/databases}\\
    2 & {connect} & {\scriptsize host: 221.5.133.18, retval: -115, por t: 8080} &                       17 & {write} & {\scriptsize /data/com.keji.danti161/shared\_prefs/com.keji.danti161.xml}\\
    22 & {write} & {\scriptsize /data/com.keji.danti160/shared\_ prefs/com.keji.danti160.xml} &         16 & {unlink} & {\scriptsize /data/com.keji.danti161/shared\_prefs/com.keji.danti161.xml.bak}\\
    21 & {unlink} & {\scriptsize /data/com.keji.danti160/shared\_ prefs/com.keji.danti160.xml.bak} &    24 & {write} & {\scriptsize /data/com.keji.danti161/databases/db.db}\\
    &          &                                                                                &    4 & {write} & \scriptsize /data/system/dropbox/drop68.tmp\\
    \bottomrule
  \end{tabular}
  \label{tbl:scal_eval}
\end{table*}

\subsubsection{Results}
The results for \emph{\textbf{a)}} and \emph{\textbf{c)}} are summarized in
Table~\ref{tbl:scal_eval_sumup} and are analyzed in the reminder of this section
with the aid of Figure~\ref{fig:scal_eval_comp_orig_rerun_1}
and~\ref{fig:scal_eval_rerun_stim} respectively. The results for
\emph{\textbf{b)}} are analyzed in depth with the aid of
Table~\ref{tbl:scal_eval} and a set of screenshots.

\mypar{ManualVsRe-exec (Figure~\ref{fig:scal_eval_comp_orig_rerun_1})} We show
the comparison between the total and distinct number of behaviors stimulated in
the original, manual test vs. the average numbers of behaviors stimulated in
re-executed tests. The rightmost plot shows the exclusive behaviors (i.e., those
stimulated only during either strategy (manual and re-execution)).

One would expect behaviors extracted in the original test to be always more than
those stimulated in re-executed tests. In some cases, this is not true, (e.g.,
in \emph{Test3}) because we are comparing behaviors exercised in different, even
if similar, applications: it is possible that an application similar to the one
originally tested generates \textbf{more} behaviors even if less stimulated. For
instance, when application \emph{A} starts it generates 10 behaviors, whereas
when application \emph{B}, similar to \emph{A}, starts it generates 20
behaviors. The same holds also for the UI stimulation, so clicking on a button
of \emph{A} we may obtain 2 behaviors, while clicking on the same button on
\emph{B} leads to 4 behaviors. Recall, however, that we are not simply counting
the exercised behaviors: In this experiment we also evaluate \emph{which}
behaviors are exclusively exercised by each strategy.

\mypar{\textbf{Re-execBehaviors (Table~\ref{tbl:scal_eval})}} A notable case is
that of a malicious behavior stimulated in the original sample, which is
exercised during the re-execution on a similar application, too. We consider the
application \texttt{com.keji.danti160}, belonging to BaseBridge malware
family. We chose this sample because during the test it showed a behavior
similar to the one shown by \texttt{com.keji.danti80}: when started, the
application asks the user to update it and installs a malicious service, named
\texttt{BridgeProvider}, on the phone. The list of behaviors extracted during
the test is presented in Table~\ref{tbl:scal_eval} (left): underlined rows
indicate the malicious actions executed by the application.

Scanning our sample repository with androsim, we found a sample, named
\texttt{com\-.keji\-.danti161} very similar to
\texttt{com\-.keji\-.danti160}. By re-executing the UI stimulation recorded with
\thesystem on the application \texttt{com\-.keji\-.danti161}, we extracted the
list of behaviors shown in Table~\ref{tbl:scal_eval} (right): underlined rows
present the same malicious actions stimulated in the original test
execution. This example illustrates that our approach can unveil behaviors
hidden to otherwise automated tests.

\mypar{AutomaticVsRe-exec (Figure~\ref{fig:scal_eval_rerun_stim})}
We now evaluate the stimulation obtained with re-execution compared with
automatic stimulation approaches. Comparing the behaviors extracted from
re-executed tests with the ones retrieved stimulating the same samples with
Monkey and \copperdroid, we obtained the data shown in
Figure~\ref{fig:scal_eval_rerun_stim}. As we can see, the re-executed
stimulation still allows to stimulate more behaviors than automatic approaches:
in fact, using \thesystem re-execution, we are able to stimulate 116\% of total
behaviors and 130\% of distinct behaviors more than automatic stimulation
methodologies. Moreover, with re-execution, we stimulate 535\% exclusive
behaviors more than Monkey and 355\% more than \copperdroid. It is also worth
noting that this is a conservative estimate of re-execution effectiveness: as a
matter of fact, our experimental data contain also cases in which re-execution
promptly failed after test beginning.

\mypar{Conclusions of Experiment~2} The results support our key intuition on the
re-execution of UI stimulation traces: we demonstrated that if (malicious)
behaviors are exercised during a manual test, it is quite likely that using the
same stimulation over the UI of similar applications will lead them to show
their behaviors during the analysis. Exercising the UI of an application with
the re-execution of stimulation traces allows to expose more behaviors than
automatic approaches.

\begin{table}[t]
  \centering
  \caption{Manual cluster analysis.}
  \scriptsize
  \begin{tabular}[t]{llll}
    \toprule
    \multicolumn{4}{c}{\emph{Random sample from 420 clusters (6,000 dataset)}}\\
    \midrule
    Classes & \#clusters (\%) & Homogeneity (\%) & Avg. Size\\
    \midrule
    1 (pure) & 84 (85.71) & 100.0 & 4.03 \\
    2        & 9 (9.1836) & 67.30 & 5.11 \\
    3        & 2 (2.0408) & 54.10 & 7.50 \\
    4        & 1 (1.0204) & 86.10 & 9.00 \\
    5        & 1 (1.0204) & 28.50 & 7.00 \\
    \midrule
    \multicolumn{4}{c}{\emph{Random sample from 628 clusters (16,000 dataset)}}\\
    \midrule
    1 (pure) & 190 (85.97)& 100.0 & 2.17 \\
    2	     & 24  (10.86) & 53.00 & 2.17 \\
    3	     & 2   (0.905)  & 53.00 & 6.50 \\
    4	     & 3   (1.358)  & 33.00 & 4.67 \\
    5	     & 1   (0.453)  & 25.00 & 8.00 \\
    6	     & 1   (0.453)  & 29.00 & 7.00 \\
    \bottomrule
  \end{tabular}
  \label{tab:quality}
\end{table}

\subsection{Experiment~3: UI Similarity}
\label{sec:ui-similarity-evaluation}
This experiments' goal is to verify that using screenshot similarity as a mean
to find apps with similar UI is a correct hypothesis. In addition, we verify
that the approach of using perceptual hashing is time and memory efficient.

\subsubsection{Dataset}
We used a first dataset of screenshots that we created by executing one app at a
time in an emulator and launching the \texttt{screencap} utility. We obtained
6,000 screenshots. This procedure took about 10 seconds per app, including the
time required to install the APK. Considering that executing a full dynamic
analysis in an instrumented environment takes time in the order of minutes, we
consider this overhead negligible. We used also a second dataset of 16,000
screenshots that we obtained by crawling the
blackmart\footnote{\url{http://www.blackmartalpha.net/}} marketplace. The
screenthots are part of each app metadata, as it happens in the majority of
marketplaces. For both the datasets we saved the images in 8-bits JPG files at
288x480 to 319x480 square pixels (at most 221KB each).

\subsubsection{Experimental setup and results}
We ran our experiments on Xeon E5506 @ 2.13GHz with 6GB of RAM. We
implemented \textbf{Phase~3} in C++ using the Boost UBLAS library, OpenMP, and
the hamming distance \verb|ph_hamming_distance| from the pHash library. In our
experiments, we used 3 concurrent OpenMP threads to calculate the sparse
distance matrix.

To verify that our approach is time and memory efficient, we executed the
\verb|ph_dct_imagehash| function to calculate the hash of each image in the our
larger dataset, which resulted in 5.030453ms on average, with a 2.172415ms
standard deviation and less than 5 megabytes of main memory on a single core. As
hashes can be indexed in proper data structures (e.g., MVP trees) that take into
account metric distances, the time required to perform a k-nearest-neighbor
search ($k = 1$) is also negligible as it grows logarithmically with the number
of apps. We verified that the time required to lookup a similar app is minuscule
with respect to the time required to run a full dynamic analysis.

\begin{figure}[t]
  \centering

  \subfloat{\includegraphics[width=.85\columnwidth]{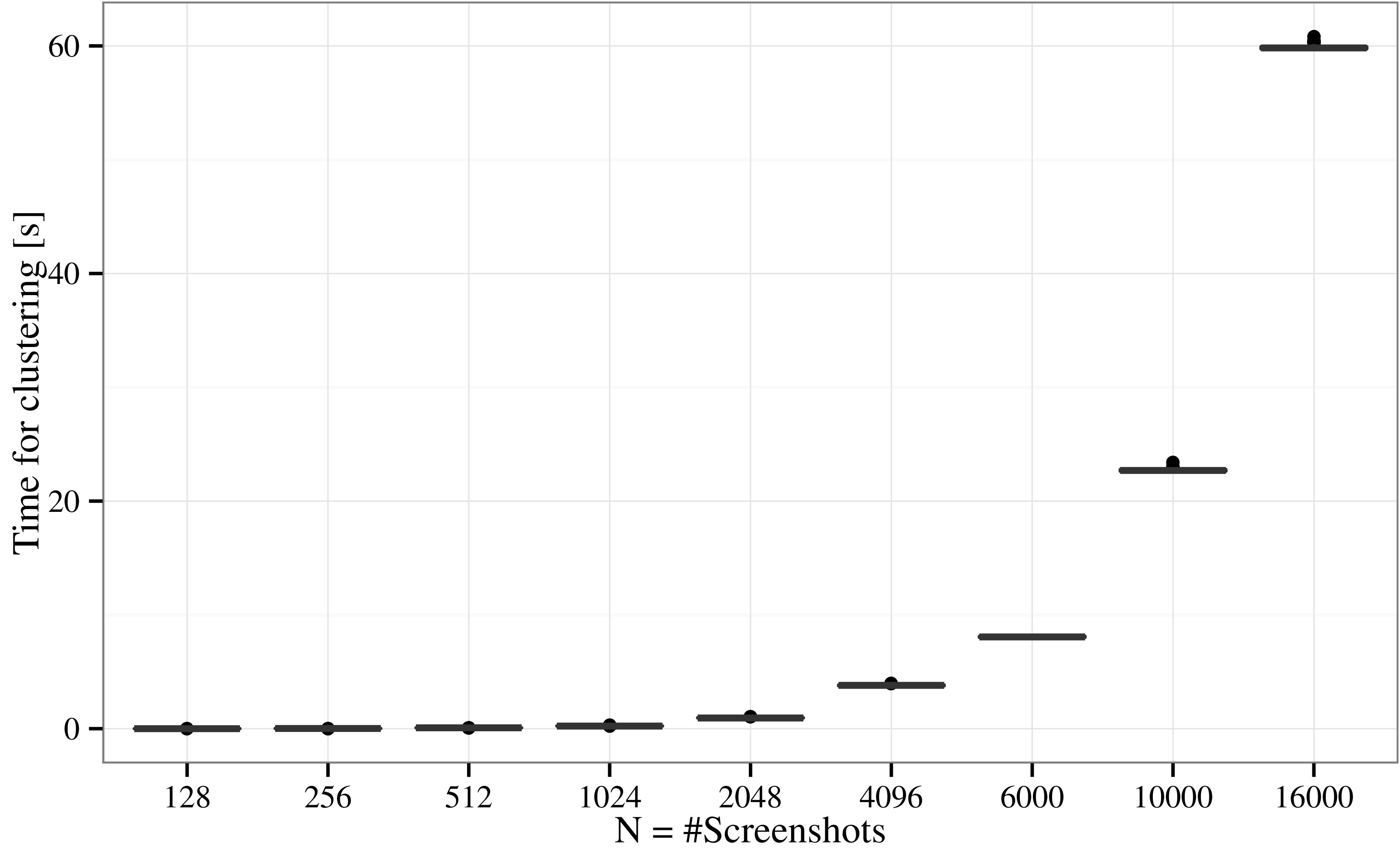}}

  \subfloat{\includegraphics[width=.85\columnwidth]{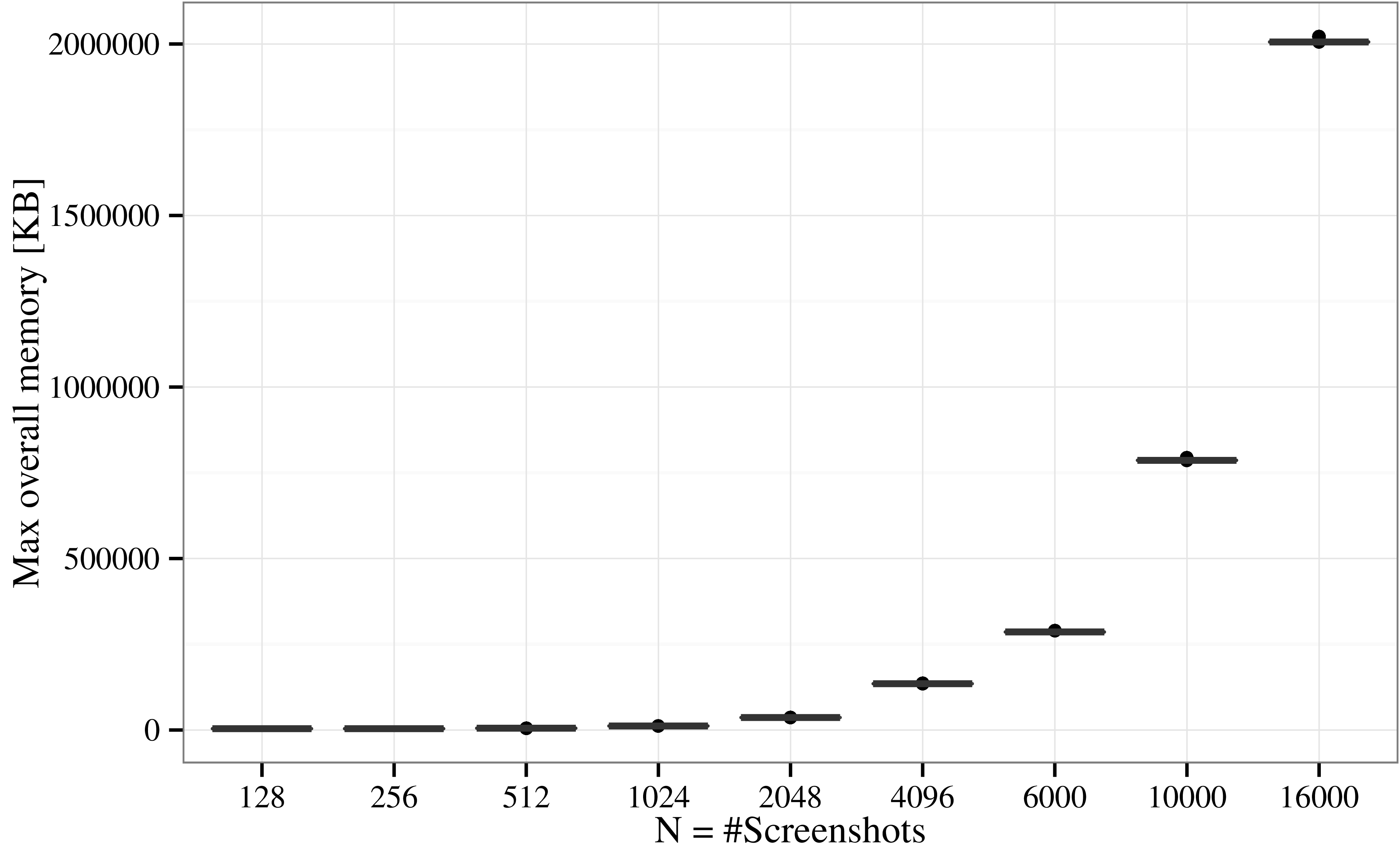}}
  \caption{Experiment 3: Boxplot distribution of time (a) and memory (b)
    requirements (33 runs per each $N$) for clustering app screenshots. Time and
    memory for comparing 2 apps is always constant.}
  \label{fig:perf_time}
\end{figure}

\begin{figure*}[t]
  \centering
  \includegraphics[width=.24\textwidth]{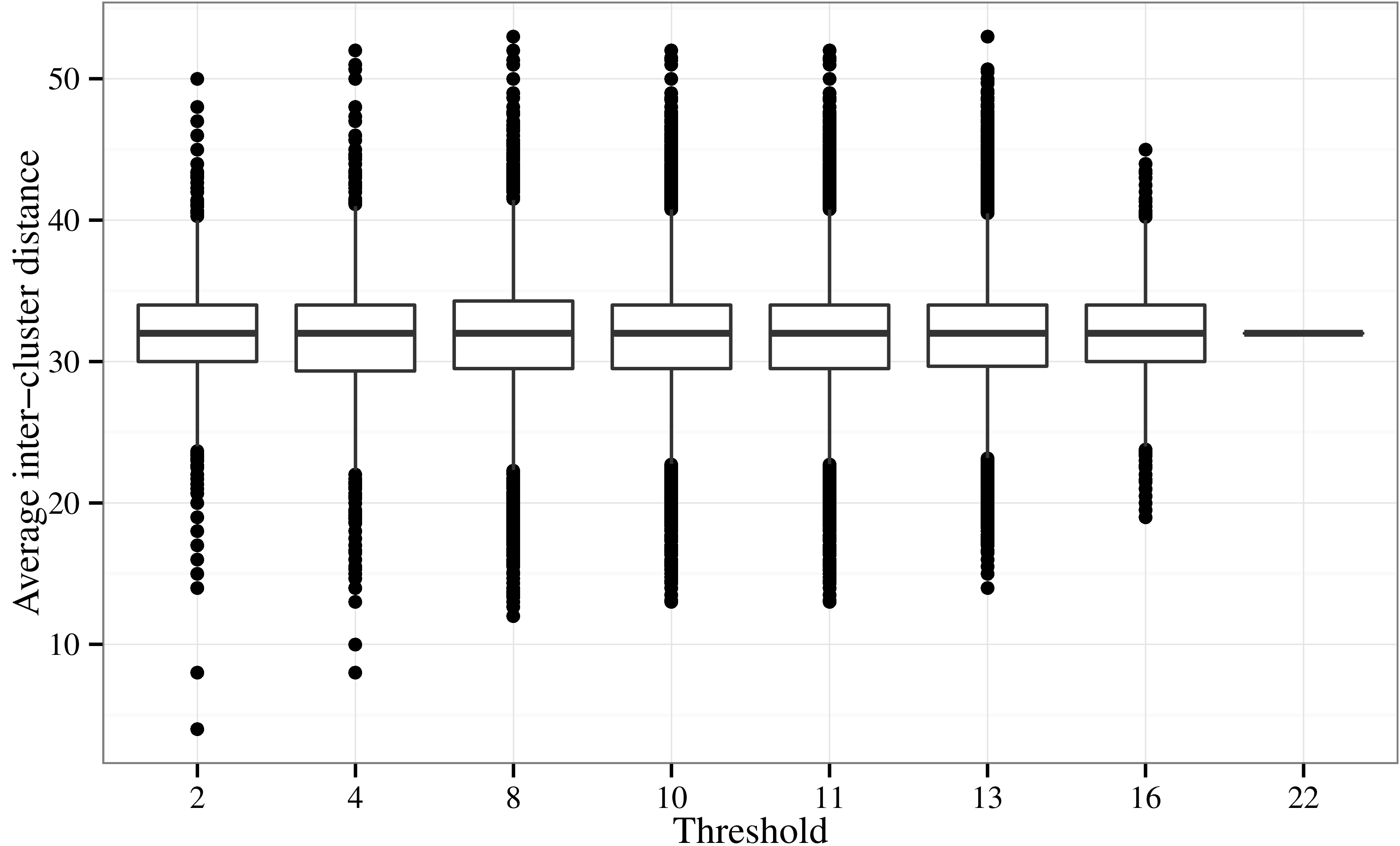}
  \includegraphics[width=.24\textwidth]{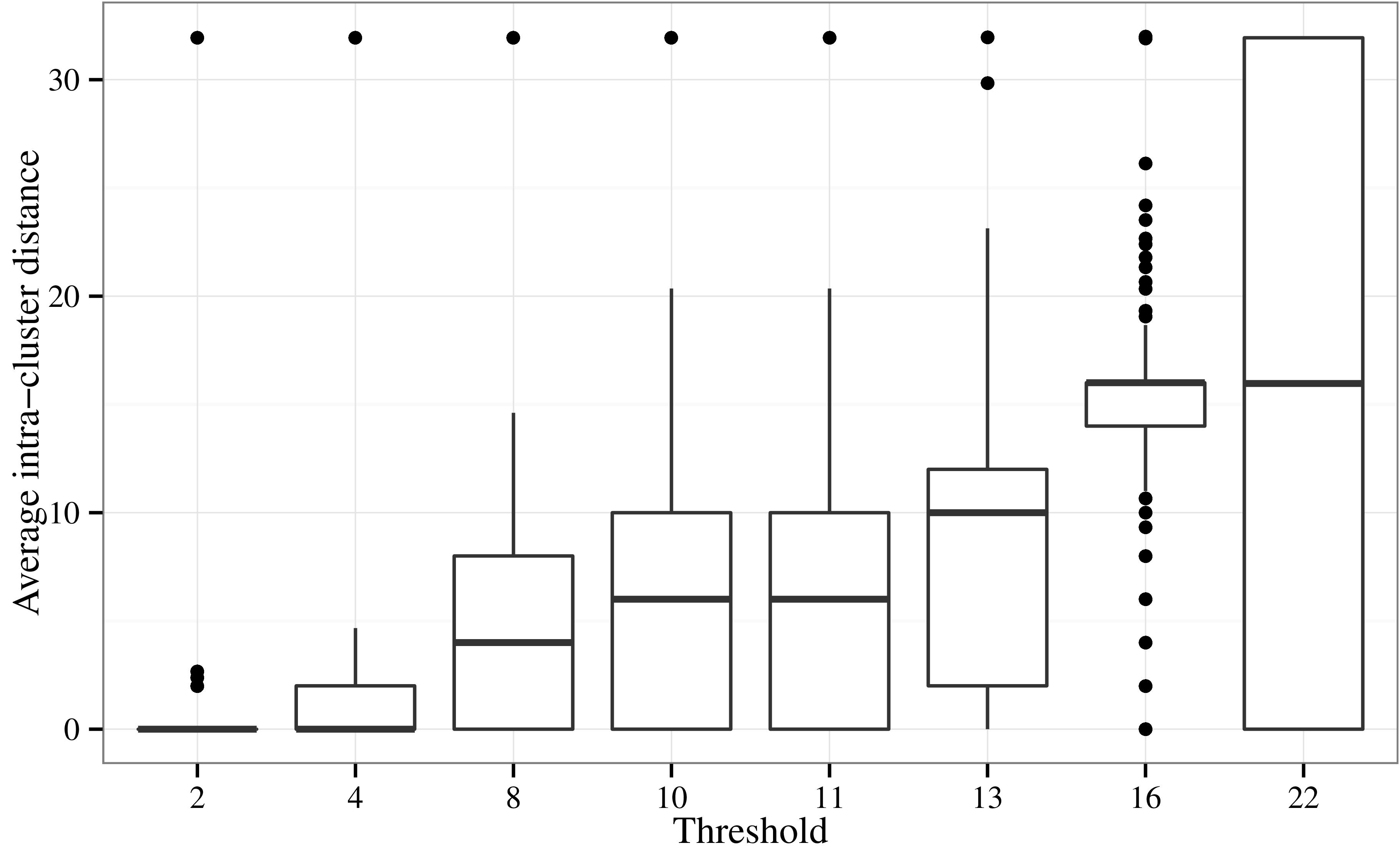}
  \includegraphics[width=.24\textwidth]{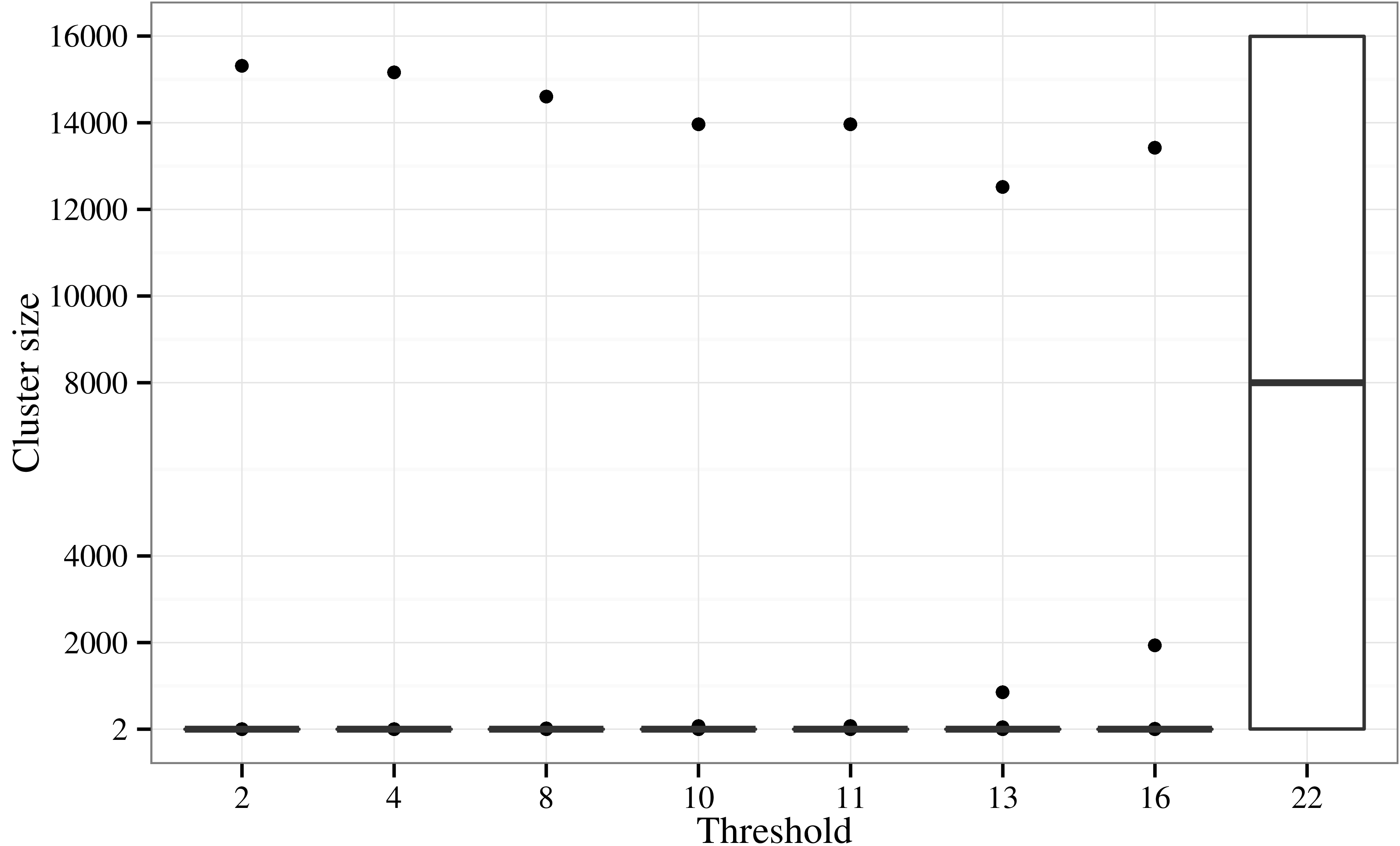}
  \includegraphics[width=.24\textwidth]{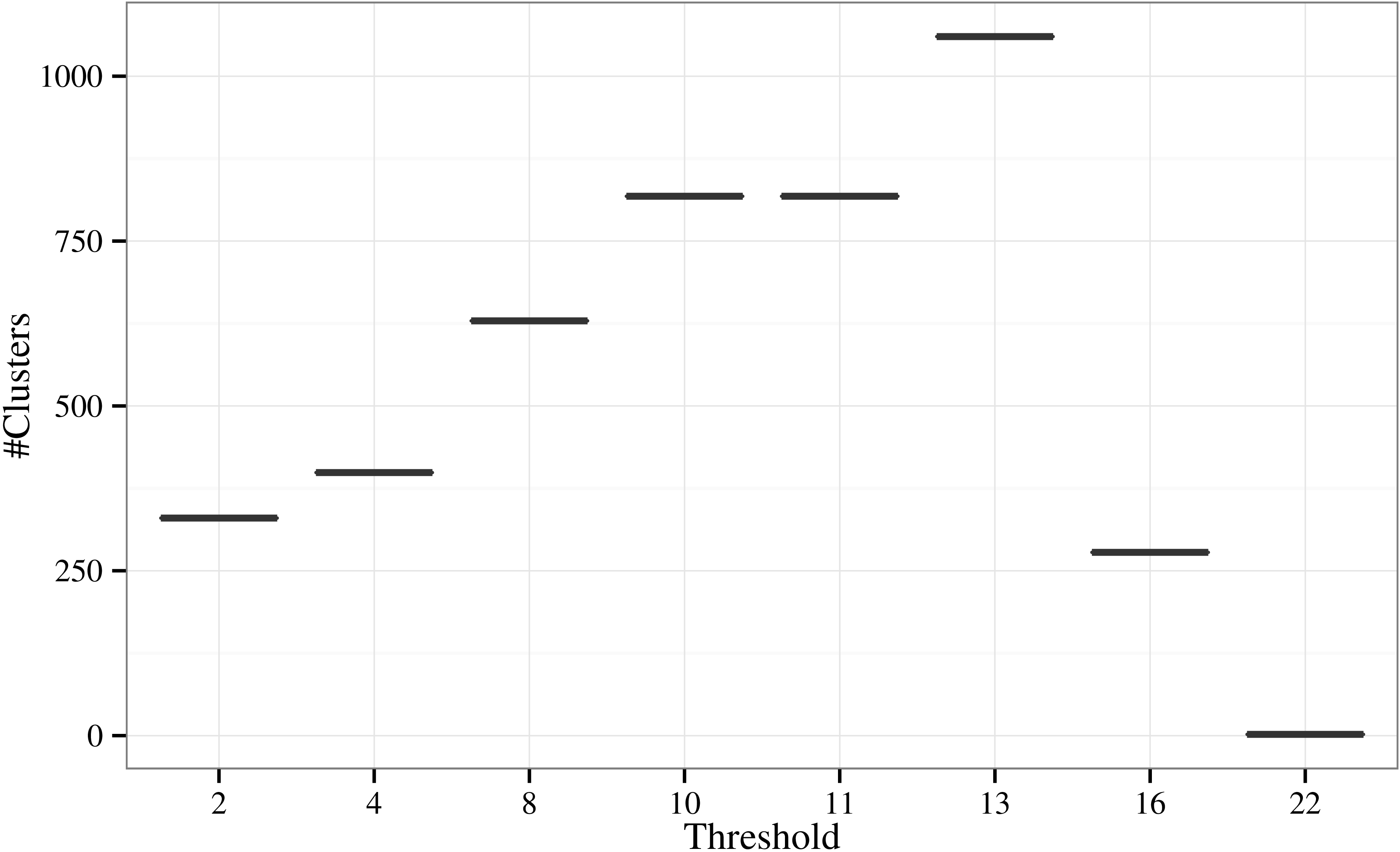}
  \caption{Experiment 3: Parameter estimation on 16,000 images. Larger versions
    in the Appendix.}
  \label{fig:parameters}
\end{figure*}

Considered the low time and memory requirements, we were able to
cluster both the datasets with DBSCAN~\cite{Ester96adensity-based}, as
demonstrated in Figure~\ref{fig:perf_time}, and allows to further speedup the
lookup phase if necessary. Moreover, in Figure~\ref{fig:parameters} we show that
the threshold on the hamming distance can be reliably chosen in an unsupervised
fashion by taking into account the intra-cluster distance, inter-cluster
distance, average cluster size and total number of clusters. We indeed observe
that increasing the threshold above 16 (bits), the number of clusters drops
significantly, while the average cluster size jumps from 2--6 elements to about
half the size of the dataset. This indicates that using a threshold below 16
allows DBSCAN to find many small clusters, each with the similar apps, plus one
noisy cluster of unpaired apps. This is also showed by the intra-cluster
distance, which increases significantly at 16. We chose 10 as the threshold, and
2 as the minimum cluster size (as we want to find, at least, couples of similar
apps).

In the Appendix we show that our approach can find interesting, non-obvious
pairs of similar apps. To validate our approach we asked Mechanical Turk Master
Qualified workers to analyze 321 randomly picked (i.e., \texttt{sort -R})
clusters and report the number and size of distinct classes of screenshots they
found in each of them. With this we calculated each cluster's homogeneity as the
size of the most frequent class over the cluster size. Ideally, homogeneity
should be 100\%, indicating 1 class of UI per cluster, which means a pure,
perfectly formed cluster. We could not reliably use the name nor the MD5 of the
application as a class label, because, as shown by previous work (see
Section~\ref{sec:related-work}), many applications are actually repackaged
versions or other applications. However, by randomly drawing 100 of 420 and 221
of 628 clusters we inspected 23.81\% and 35.19\% of the clusters respectively
from the 6,000 and 16,000-images datasets. As Table~\ref{tab:quality} shows, our
approach finds 85.71--85.97\% pure clusters), whereas the reminders have a
reasonably high homogeneity, except for some outliers.

%% file: future.tex
\section{Limitations and future work}
\label{sec:limitations}
The presence of a very small view object in the original sample layout that is
not present in the layouts of similar applications is a corner case that can
make our re-execution incomplete (see Appendix). For example, during the
stimulation of the original sample, our user clicked on a link embedded in a
TextView object. Re-executing the test on a similar application, the content of
the TextView changed, with consequent vanishing of the link. Hence, clicking on
the TextView in the original sample led to open a new window, while the same
click on the similar application did not generate any transition, making the
re-execution test fail.

To avoid this specific cases of dynamic layouts, our future work includes
attaching semantic tags to each screenshot (e.g., list of known view objects
visualized), so as to devise a similarity criterion that can recognize whether
two layouts are very similar, yet with a significant tiny variation (e.g,
absence of a single, small button). However, this creates the further challenge
of deciding a threshold, because such a semantic similarity criterion cannot
possibly be mapped on a metric space. Instead, our current method is simple and
practical because it requires no threshold: We find the application that
minimizes the distance, and we can do this because the features are metric.


%% file: related.tex
\section{Related work}
\label{sec:related-work}
Our work is related to dynamic analysis, similarity and UI exercising of Android
applications.

\mypar{Dynamic analysis} TaintDroid~\cite{taintdroid}, integrated and extended
by other analysis systems such as DroidBox~\cite{droidbox} and
Andrubis~\cite{andrubis_blog}, extends Android to taint track privacy-sensitive
resources and notify the user if such information leave the system via network,
SMS, or else. Unfortunately, it prevents third-party apps from loading native
libraries, and is version specific. DroidBox~\cite{droidbox} extends TaintDroid
with the ability to keep track of network traffic, sent SMS, phone calls,
etc. DroidBox has recently been upgraded to APIMonitor, which works directly on
the source code of applications (rather than on the code of the system).

Differently from previous approaches, \copperdroid{}~\cite{copperdroid}, is an
out-of-the-box dynamic analysis tool that relies on an instrumented version of
the Android emulator to automatically reconstruct behaviors. By thoroughly
inspecting syscalls and their arguments, \copperdroid{} performs an unified
analysis of both low level (e.g., file writes) and high level (e.g., send an
SMS) actions performed by an application. Furthermore, it uses an effective
stimulation mechanism that increases the code coverage. Similarly, DroidScope is
built on top of QEMU. Authors modified the translation phase from Android code
to TCG, an intermediate representation used in QEMU, to insert extra
instructions that enable fine-grained analyses. To reconstruct the two
semantic levels of Android, (Linux and Dalvik), VMI is leveraged.

\mypar{Application Similarity} Tools such as
Androguard\footnote{\url{https://code.google.com/p/androguard/}} assist reverse
engineers in finding similar APKs, but have accuracy and scalability
issues. Therefore, research in this direction is fairly active for different
purposes. For example,~\cite{gibler2013adrob,Chen:2013:UEA:2484417.2484419} use
app similarity to find repackaged, ad-aggressive versions of applications
distributed on alternative markets.

Juxtapp~\cite{hanna12:_juxtap} recognizes whether applications contain
known, flawed code, exhibit code reuse that indicates piracy, or are
(repackaged) variants of known malicious apps. Juxtapp focuses on
scalability, proposing a similarity metric that is suitable for
map-reduce frameworks. Juxtapp requires 100 minutes of computation on
100 8-core machines with 64GB of RAM to analyze 95,000 distinct
APKs. DNADroid~\cite{Crussell:2012:DNADroid} exploits the dependency
graph to find pairs of matching methods to recognize plagiarized
applications.

\thesystem differs substantially from previous work because it takes the UI into
account. Our goal is not that of finding similar \emph{code}, but to finding
similar \emph{interfaces}.

\mypar{Exercising of Android applications} Dynodroid~\cite{dynodroid} uses an
\emph{``observe and execute''} approach (i.e., analyze the content of displayed
UI elements and then generate tailored random input events). Dynodroid 
reaches the same code coverage obtained with Monkey, but with much less events.

SmartDroid~\cite{smartdroid} leverages static and dynamic analysis to extract a
sequence of UI events that allow to stimulate suspicious behaviors. Static
analysis is used to identify the invocations of sensitive methods. Then,
sensitive paths from application's entry points to identified method invocations
are built. Last, dynamic analysis is used to verify the validity of the paths
previously found.

ACTEve~\cite{anand2012automated} proposes an algorithm that leverages concolic
execution~\cite{sen2005cute} to automatically generate input events. It uses
advanced subsumption and pruning algorithms to avoid the path explosion problem
and, thus, is able to automatically generate test inputs that strive the
execution flow of an application to get high code coverage. Despite its
optimization, though, the overhead of this technique is high (i.e., hours) for
malware analysis purposes.

Finally, RERAN~\cite{reran} allows to record and replay low-level UI events
directly reading from, and writing on, system input devices. This work uses an
approach similar to the one used by \thesystem to inject input events, but it is
limited to a mere re-execution of the original recorded touch events without
offering any analysis of application UI.

\thesystem differs substantially from previous work because (1) takes human
users into account, (2) introduces the concept of visual similarity and, at the
same time, (3) binds low-level input events to view objects.


%% file: conclusion.tex
\section{Conclusion}
Dynamic analysis is facing new challenges with mobile malware. Mobile
software (both goodware and malware) was born in a radically different
ecosystem than traditional software, which includes, for instance, app
marketplaces---the main distribution channel for malicious apps.

Because of this different nature, the victim's participation during the
infection is essential, and greater than in traditional malware. We believe that
orthogonal approaches to dynamic analysis, such as \thesystem, that strive to
capture the user's actions, are an important research direction to pursue. Our
experiments show that our hypotheses are true, and that human users can be
effectively and efficiently included in the dynamic analysis workflow, also
thanks to the availability and accessibility of crowdsourcing platforms.

This can potentially change the way we conduct dynamic analysis of mobile
applications (from fully automatic, to scalable and collaborative): We believe
that our system can attract the interest not only of security analysts but also
of normal users that want to safely test potentially malicious applications.


%% file: appendix.tex
\appendix

\subsection*{Sample Output of Phase~3}
We show 6 sample clusters created by our approach, which highlight how it can
find non obvious UI-similar applications.
\smallskip
\includegraphics{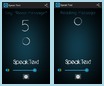}
\includegraphics{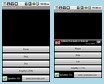}

\includegraphics{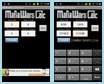}
\includegraphics{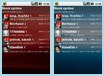}

\includegraphics{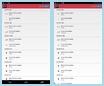}
\includegraphics{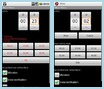}

\subsection*{Larger Version of Figure~\ref{fig:parameters}}
\begin{center}
  \includegraphics[width=.7\columnwidth]{pics/experiment3/speed-parameters/16k/th_avg_inter.png}
  \includegraphics[width=.7\columnwidth]{pics/experiment3/speed-parameters/16k/th_avg_intra.png}
  \includegraphics[width=.7\columnwidth]{pics/experiment3/speed-parameters/16k/th_cluster_size.png}
  \includegraphics[width=.7\columnwidth]{pics/experiment3/speed-parameters/16k/th_clusters.png}
\end{center}

\subsection*{Sample Corner Case in Phase~2}
Example of re-execution failure due to the presence of particular UI
elements. See Section~\ref{sec:limitations}.
\begin{center}
  \includegraphics[width=0.8\columnwidth]{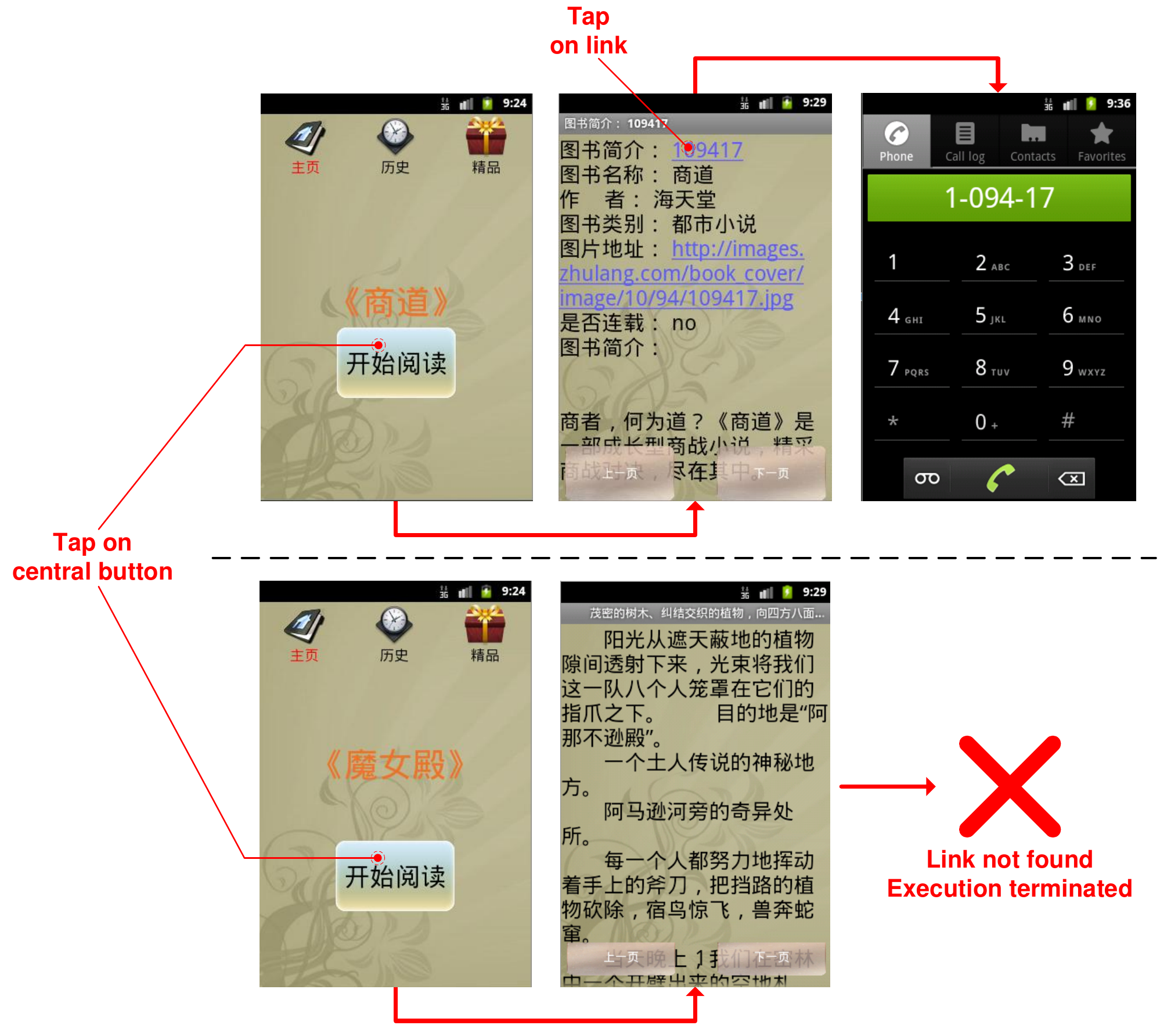}
\end{center}
